\documentclass[prd,twocolumn,showpacs,nofootinbib]{revtex4}
\usepackage{bm}
\usepackage{graphicx}
\usepackage{amsmath}
\usepackage{wasysym}
\begin{document}
\def \lu {Lundberg \& Edsj{\"o}~}
\newcommand{\mwimp}{$m_\chi$}
\newcommand{\sigmapsi}{$\sigma_p^{SI}$}
\newcommand{\sigmapsd}{$\sigma_p^{SD}$}
\newcommand{\sigmansd}{$\sigma_n^{SD}$}
\newcommand{\kms}{km~s$^{-1}$}
\newcommand{\vsun}{$v_\odot$}
\newcommand{\ve}{$v_\oplus$}
\newcommand{\vescsun}{$v^{\odot}_{esc}$}
\newcommand{\vesce}{$v^{\oplus}_{esc}$}

\title{Dark matter in the solar system III: The distribution function of WIMPs at the Earth from gravitational capture.}
\author{Annika H. G. Peter}
\email{apeter@astro.caltech.edu}
\affiliation{Department of Physics, Princeton University, Princeton, NJ 08544, USA}
\affiliation{California Institute of Technology, Mail Code 105-24, Pasadena, CA 91125, USA}
\date{\today}

\begin{abstract}
In this last paper in a series of three on weakly interacting massive particle (WIMP) dark matter in the solar system, we focus on WIMPs bound to the system by gravitationally scattering off of planets.  We present simulations of WIMP orbits in a toy solar system consisting of only the Sun and Jupiter.  As previous work suggested, we find that the density of gravitationally captured WIMPs at the Earth is small and largely insensitive to the details of elastic scattering in the Sun.  However, we find that the density of gravitationally captured WIMPs may be affected by external Galactic gravitational fields.  If such fields are unimportant, the density of gravitationally captured WIMPs at the Earth should be similar to the maximum density of WIMPs captured in the solar system by elastic scattering in the Sun.  Using standard assumptions about the halo WIMP distribution function, we find that the gravitationally captured WIMPs contribute negligibly to direct detection event rates.  While these WIMPs do dominate the annihilation rate of WIMPs in the Earth, the resulting event rate in neutrino telescopes is too low to be observed in next-generation neutrino telescopes.
\end{abstract}


\maketitle

\section{Introduction}\label{sec:intro}

\subsection{Dark Matter Detection in the Solar System}\label{sec:intro_dm}
A number of lines of evidence point to the existence of a significant amount of dark matter in the universe, although its identity is a mystery \cite[e.g.,][and references therein]{bertone2005,komatsu2008}.  Perhaps the most popular candidate for dark matter is a species of WIMP, both because such particles appear naturally in extensions to the Standard Model of particle physics and because such particles have astrophysically desired properties \cite{jungman1996,cheng2002,hubisz2005}.

There are experiments underway to identify WIMPs in collider experiments (e.g., the LHC \cite{arkanihamed2006,baltz2006,hooper2008a}) or by their annihilation productions throughout the Galaxy \cite{wai2007,pamela2008,chang2008}, although it will be challenging to conclusively determine WIMP properties.  In the case of collider experiments, WIMPs cannot be directly observed, and mapping any anomalous interactions to a particular theory will be difficult.  Astrophysical detection of WIMPs is complicated by both the uncertainty in the distribution of dark matter in the densest parts of the Galaxy as well as poorly characterized foregrounds \cite[e.g.,][and references therein]{aharonian2005,bertone2005,aharonian2006,zaharijas2006,profumo2008}.

There are several experiments to look for WIMPs in the solar system.  Direct detection experiments look for the tiny ($\sim 10-100\hbox{ keV}$) recoils of nuclei struck by WIMPs.  Current experiments have $\sim 10\hbox{ kg}$ of fiducial target mass.  The XENON10 \cite{angle2008,angle2008b} and CDMS \cite{cdms2008} experiments currently have the best constraints on the spin-independent WIMP-proton $\sigma_p^{SI}$ and spin-dependent WIMP-neutron $\sigma_n^{SD}$ WIMP-neutron elastic scattering cross sections, at the levels of $\sigma_p^{SI} \lesssim 4\times 10^{-44}\hbox{ cm}^2$ and $\sigma_n^{SD} \lesssim 10^{-38}\hbox{ cm}^2$ for WIMP mass $m_\chi \sim 100\hbox{ GeV}$.  Upcoming experiments should gain a factor of $\sim 100$ in sensitivity \cite{aprile2002,akerib2006c,hime2007,gaitskell2007}.

Neutrino telescopes are searching for neutrinos from WIMP annihilation in the cores of the Earth and the Sun.  The current best constraint on the WIMP-proton elastic scattering cross section $\sigma_p^{SD} \lesssim 10^{-39}\hbox{ cm}^2$ for $m_\chi \sim 100\hbox{ GeV}$ come from flux limits of neutrinos from the Sun \cite{desai2004,ackermann2006}.  

While the backgrounds in direct detection experiments and neutrino telescopes have been meticulously examined, the astrophysical properties of WIMPs need to be understood either in order to make accurate predictions for the event rates in such experiments, or to constrain particle physics models from data.  Thus, the distribution function (DF) of WIMPs in the solar system needs to be characterized.  Aside from uncertainties in the DF of halo WIMPs streaming through the \cite[e.g.][]{bergstrom1998b,helmi2002,read2008}, there is theoretical uncertainty in the DF of WIMPs bound to the solar system.  As has been demonstrated by several authors \cite{damour1999,bergstrom1999,gould2001,lundberg2004}, this latter population may have a profound impact on predictions for direct detection event rates and on the annihilation rate of WIMPs captured in the Earth.

In order to characterize the bound WIMP population, we have undertaken a program of simulating WIMP orbits in the solar system, taking into account the possibility of further scattering in the Sun.  The results from our simulations in a toy solar system consisting of Jupiter on a circular orbit about the Sun are described in a series of three papers, of which this is the last.  In Paper I \cite{peter2009a}, we simulated a population of WIMPs bound to the solar system by elastic scattering in the Sun, a population originally postulated by \citet{damour1999}.  We found that the DF of this population depended on both the WIMP mass and the strength of the WIMP-baryon interaction, but that the population was too small to significantly enhance direct detection event rates or to produce an observable signature of WIMP annihilation in the Earth.  However, in Paper II of the series \cite{peter2009b}, we found that the extended lifetime of WIMPs captured in the Sun had interesting consequences for searches for WIMP annihilation in the Sun.  Both the gravitational interactions between WIMPs and planets and finite optical depth in the Sun to WIMPs altered the standard picture that all WIMPs captured in the Sun immediately thermalize.  These modifications to the standard thermalization pictured imply lower annihilation rates of WIMPs in the Sun for $m_\chi \gtrsim 1$ TeV or for low elastic scattering cross sections.

\subsection{Gravitationally Captured WIMPs}\label{sec:intro_grav}
In this paper, Paper III, we present simulations of the orbits of another class of bound WIMPs: those captured in the solar system by gravitational interactions with the planets.  Previous work on this population has relied on treating all WIMP-planet encounters as local, and with only a crude treatment of WIMP-baryon encounters in the Sun.  In order to make comparisons with our simulations, we briefly outline the previous work on the gravitationally bound WIMP population.

The first to estimate the size of the gravitationally bound WIMP population was Gould \cite{gould1988,gould1991}.  Gould approximated all encounters between planets and WIMPs as local; gravitational scatters do not change the WIMP speed with respect to the planet, but do change the orientation of the orbit.  This translates to a change in velocity in the heliocentric frame, and hence, to changes in the WIMP energy and angular momentum.  Using a random walk approach, Gould found the average time for the angle between the direction of WIMP velocity with respect to the planet and the direction of motion of the planet to change by order unity as a function of WIMP speed, the ``angular diffusion'' timescale.  He interpreted this as the timescale on which the planets could move WIMPs in or out of a particular region of phase space, since whether a WIMP is bound or unbound to the solar system depends on the direction of the WIMP velocity in a planet-centric frame with respect to the planet motion.  Gould then estimated the DF of WIMPs at the Earth using the following detailed balance approximation.

Assuming that Galactic halo WIMPs can be treated as having a Maxwellian DF near the solar system, and that the Galactic dark matter halo is non-rotating in an inertial Galactocentric frame, the DF of low speed WIMPs (ones that may be captured gravitationally) is nearly constant, $f(v) \approx f$.  Since the escape speed from the solar system at the position of the planets is small, the phase space density of halo WIMPs at each planet is also $\approx f$.  Gould argued that if the angular diffusion timescale were less than the age of the solar system, the phase space density of bound orbits should be the same as the phase space density of the unbound halo WIMPs for a given WIMP speed in an inertial frame moving with the planet.  The flow of WIMPs filling the bound phase space is countered by the flow of WIMPs becoming unbound to the solar system.  The angular diffusion timescale associated with Jupiter is of order Myr for any part of phase space accessible to Jupiter.  In Gould's picture, the phase space corresponding to bound Jupiter-crossing WIMPs should have the same density as the phase space associated with unbound orbits.  Furthermore, Gould found that the timescale associated with the Earth and Venus for speeds with respect to the Earth of $u < 30 \hbox{ km s}^{-1}$ was less than the age of the solar system.  Thus, for such speeds, the WIMP phase space density should be the same for any orientation of the velocity vector in an inertial frame moving with the Earth (geocentric).  Some parts of phase space for $u > 30\hbox{ km s}^{-1}$ is empty in this picture, but the majority of the accessible phase space at those speeds corresponds to unbound or Jupiter-crossing orbits.  Hence, the speed distribution of WIMPs at the Earth in a frame moving with the Earth should be identical to the speed distribution of halo WIMPs in free space (outside the potential well of the Sun) for $u < 30 \hbox{ km s}^{-1}$.  Gould found that the free space approximation was reasonable for larger speeds, too.

Gould neglected WIMP-nucleus scattering in the Sun.  To determine the importance of this effect, \citet{lundberg2004} also treated WIMP-planet encounters as local, and solved a gravitational diffusion equation for WIMP orbits in a solar system consisting of Jupiter, the Earth, and Venus.  The Sun was either treated as a point mass or as infinitely optically thick to WIMPs.  The timescale for hitting the Sun was estimated using a small set of individual WIMP orbit simulations.  They found that the DF for WIMPs if the Sun were infinitely optically thick to WIMPs was substantially smaller than if the Sun were a point mass.  

\subsection{This Work}\label{sec:intro_work}
In light of previous work on gravitationally captured WIMPs, there are several reasons to perform suites of WIMP orbit simulations.  First, the work of \citet{lundberg2004} suggests that scattering in the Sun is an important loss mechanism for bound WIMPs.  It would be useful to understand the degree of depletion as a function of the WIMP optical depth in the Sun.  Second, both Gould and Lundberg \& Edsj{\" o} treat WIMP-planet interactions as local.  However, these treatments neglect long-range encounters, short-period interactions, and is fundamentally insensitive to resonances (although these are incorporated for a set of Earth-crossing WIMPs in \lu), which are known to be important in determining the dynamics of the population of WIMPs bound to the solar system by elastic scattering in the Sun (Paper I) and of minor bodies in the solar system \cite{wisdom1982,farinella1994,duncan1997,gladman2000}.  Finally, since WIMPs tend to be captured on initially very loosely bound orbits, they may be affected by external gravitational fields, which are known to be important in shaping the Oort cloud \cite{heisler1986,duncan1987}.

In this work, we present simulations the gravitational capture and evolution of WIMPs in the solar system.  As in Paper I, we simulate orbits in a toy model solar system consisting of Jupiter on a circular orbit about the Sun.  The reasons for using a simplified system are twofold. (i) Since the integration algorithm is new, it is useful to check its performance in a simple system.  The toy system we use has a constant of motion, the Jacobi constant; its constancy throughout the simulations was an indication of the accuracy of the integration scheme.  (ii) Since Jupiter is by far the largest planet in the solar system, it should dominate the dynamics of WIMPs.  These simulations provide a solid basis for understanding the dynamics of WIMPs in more complicated systems, which we hope to simulate in the future.

We describe the simulations in Section \ref{sec:simulations}, and present the resulting DFs in Section \ref{sec:df}, which we construct using a method outlined in Appendix \ref{sec:df_estimator}.  In that section, we also discuss the DF in context of solar depletion and Galactic gravitational fields.  We show the direct detection and neutrino telescope event rates from the gravitationally bound WIMP population in Sections \ref{sec:dd} and \ref{sec:id}.  In each of those sections, we compare the event rates from gravitationally captured WIMPs to the population of WIMPs bound to the solar system by elastic scattering in the Sun, the subject of Paper I.  In Section \ref{sec:discussion}, we discuss the our results in context of other work.  The key points of this work are summarized in Section \ref{sec:conclusion}.

\section{Simulations}\label{sec:simulations}

Simulations were performed using the algorithm described in Paper I, which we briefly summarize here.

For most of the WIMP path, we integrated the orbits using a symplectic integrator optimized for systems in which one body dominates the gravitational potential of the system, and for which the gravitational force does not deviate significantly from an inverse square law \cite{mikkola1999,preto1999}.  Symplectic integration is desired for long-term orbit integrations because errors are oscillatory instead of growing with time.  This particular symplectic integrator is efficient for integrating the highly eccentric orbits characteristic of WIMPs because it allow for variable time steps (short at perihelion, long near aphelion).  To achieve this in a symplectic way, it treats time $t$ and the WIMP energy $-p_0$ as conjugate variables; each step in time $\Delta t$ is related to a step in the new fictitious time coordinate $\Delta s$ by
\begin{eqnarray}
	\Delta t = g(\mathbf{r},\mathbf{p}, t) \Delta s,
\end{eqnarray}
where $\mathbf{r}$ and $\mathbf{p}$ are the WIMP position and momentum coordinates, respectively.  For the choice
\begin{eqnarray}
	g(\mathbf{r},\mathbf{p}) = - \frac{GM_\odot}{\Phi(\mathbf{r},t)},\label{eq:g}
\end{eqnarray}
where $\Phi(\mathbf{r},t)$ is the gravitational potential, \citet{preto1999} show that the integrator exactly traces the solution to the two-body problem with only a phase error.  Since typically $\Phi \approx -GM_\odot / |\mathbf{r} - \mathbf{r}_\odot|$, where $\mathbf{r}_\odot$ is the position of the Sun, $\Delta t \propto | \mathbf{r} - \mathbf{r}_\odot|$, so that for fixed $\Delta s$, shorter time steps are taken at perihelion than aphelion.

We integrate the eight-dimensional equations of motion using a second-order leapfrog mapping using a fixed fictitious time step $\Delta s = h$.  We use only a second order integrator because we are interested in the behavior of an ensemble of orbits, and not the precise orbits of individual WIMPs.  Even for this low order mapping, there is no numerical precession of orbits, and errors in the Hamiltonian are oscillatory in nature.  

Although it would be ideal use the symplectic integrator with fixed fictitious time step $h$ throughout the integrations, it is too time consuming to be practical.  This is because the integrator is not optimized to handle potentials that deviate significantly from $\Phi(\mathbf{r},t) = -GM_\odot/|\mathbf{r} - \mathbf{r}_\odot |$, which is the case in the interior of the Sun or when WIMPs experience close encounters with planets.  In those cases, $h$ would need to be set prohibitively small in order to resolve those potentials.

Instead, we treat passages through the Sun and close encounters with planets using alternate methods, which allows for $h$ to be set to a reasonably large value.  While this breaks the Hamiltonian flow of the symplectic scheme, we have taken care to insure that our methods for treating the special cases minimize errors in the Hamiltonian.

For the passages through the Sun, we exploit the fact that tidal forces from the planets are much smaller near the Sun than they are elsewhere in the orbit.  We treat passages through the Sun as a two-body problem, and are able to map the coordinates of the WIMP as it enters the Sun to its coordinates upon exit.  In Paper I, we show that this method does not induce additional numerical errors.  

We define a sphere (or ``bubble'') around each planet in which we allow another break to the symplectic algorithm described above.  In this bubble, we still integrate the WIMP orbits using the symplectic algorithm, but with a new fictitious time step $h^\prime$ tuned to achieve a minimum accuracy criterion for the Hamiltonian. To find this fictitious time step, we first integrate the WIMP orbit with the fiducial $h$ for the region outside the planet bubble and outside the Sun.  If a minimum accuracy requirement $ \Delta E / E = | p_0 + E(\mathbf{r},t) | / E(\mathbf{r},t)$ is met, then the orbit is allowed to continue.  If the minimum accuracy requirement is not met, the trajectory through the bubble is integrated with a slightly smaller $h^\prime$.  This procedure is iterated until the minimum accuracy requirement is met or the error plateaus.  We tune the bubble size, the minimum accuracy requirement and $h$ to minimize the overall integration time while maintaining small oscillatory errors in the Hamiltonian throughout the integration.  In later sections, we describe our specific choices for these variables.

\subsection{(Astro)Physical Assumptions}\label{sec:ss_model}

\emph{The Solar System:}  We perform simulations of WIMP orbits in a toy solar system consisting of Jupiter (\jupiter) on a circular orbit of $a_{\text{\jupiter}} = 5.203\hbox{ AU}$ the Sun (modeled using \cite{bahcall2005}).  This system admits a constant of motion, the Jacobi constant, which is a useful check on the accuracy of the integration algorithm.  For simplicity, we treat Jupiter as having a constant mass density; this is not a realistic representation of Jupiter's structure, but only a tiny percentage of simulated WIMPs ever go through the planet.  WIMP-baryon encounters in Jupiter are also neglected for similar reasons; in addition, the optical depth of Jupiter to WIMPs is negligible compared to that in the Sun.  Since the kinetic energy and speeds of solar nuclei are small compared to those of the WIMPs, we treat solar nuclei as being at rest with respect to the Sun.

\emph{Dark Matter:}  The scattering probability of WIMPs in the Sun is completely determined by the solar model, the WIMP mass $m_\chi$, and the cross sections $\sigma_p^{SD}$ and $\sigma_p^{SI}$.  Since we suspected that the bound WIMP DF would not strongly depend on scattering in the Sun, we chose only one point in the WIMP mass-cross section parameter space to use for the simulations, $m_\chi = 500$ AMU, $\sigma_p^{SD} = 0$, and $\sigma_p^{SI} = 10^{-43}\hbox{ cm}^2$.  This point lies below the best limits on WIMP parameter space from direct detection experiments using standard assumptions about the halo WIMP DF \cite{angle2008,cdms2008}.  However, in order to extrapolate our DFs to other points in WIMP parameter space, we kept track of the integrated optical depth of each WIMP as a function of time.

\subsection{Starting Conditions}\label{sec:start}

In deciding how to arrange the initial conditions, it is useful to think about the flux of dark matter particles into a sphere of radius $R$ centered on the Sun.  The flux for an isotropic distribution function $f$ is
\begin{eqnarray}
	F( R, v ) &=&  4 \pi v^2 f( v_s(R,v) ) \times \frac{1}{2} v \cos \theta \text{d} \cos \theta \\
	&=& \pi v^3 f(v_s(R,v)) \text{d} v \text{d}(\cos^2 \theta),
\end{eqnarray}
where $\pi/2 < \theta < \pi$ is the angle between the velocity $\mathbf{v}$ and the position vector $\mathbf{R}$ for incoming particles, and $v_s$ is the speed of the particle relative to the Sun but far outside its gravitational sphere of influence.  We have invoked Liouville's theorem to find the WIMP phase space density at an arbitrary distance from the Sun.  The total number of particles going inward through this spherical shell per unit time is
\begin{eqnarray}
	\dot{N}(R) = 4 \pi^2 R^2 v^3 f( v_s(R,v)) \text{d}v \text{d}(\cos^2 \theta).
\end{eqnarray}
It is useful to express this rate in terms of the specific energy $E$ and specific angular momentum $J$ instead of $v$ and $\cos^2 \theta$.  Given that
\begin{eqnarray}
	E &=& \frac{1}{2} v^2 + \Phi_\odot (R) \\
	J &=& Rv \sin \theta,
\end{eqnarray}
We find
\begin{eqnarray}
	\dot{N} = \pi f\left( \sqrt{2 E} \right) \text{d} E \text{d} J^2. \label{eq:gravflux}
\end{eqnarray}
Therefore, the number of particles going through any shell is independent of the radius of the shell for a given energy and angular momentum; this is to be expected since there is no loss of particles between shells.
\newline\indent
If we were to sample all particles that flow in towards the Sun, we would sample the energy according to to $f(\sqrt{2E})$ and the angular momentum to be uniform in $J^2$.  However, by restricting the range of incoming particles that are sampled to those that could be scattered onto bound orbits, we can speed up the calculation. 

To find the range of $E$ for which particles might possibly be gravitationally scattered by Jupiter onto bound orbits, it is useful to think of gravitational capture in the following way.  In the frame of the planet, the particle speed does not change during the encounter, but its direction with respect to the direction of motion of the planet does.  If the particle has a velocity $\mathbf{v}$ with respect to the Sun before encountering Jupiter, it will have an initial speed with respect to Jupiter of $\mathbf{u} = \mathbf{v} - \mathbf{v}_{\text{\jupiter}}$, where $\mathbf{v}_{\text{\jupiter}}$ is the velocity of Jupiter with respect to the Sun. After encountering Jupiter, the particle will have a velocity $\mathbf{u}^\prime$ with respect to Jupiter and $\mathbf{v}^\prime = \mathbf{u}^\prime + \mathbf{v}_{\text{\jupiter}}$ with respect to the Sun.  For particles that were barely unbound to the solar system to begin with, it takes only a tiny deflection of the orbit to bind it to the solar system.  However, for particles with increasingly higher energy with respect to the Sun, it takes an ever greater deflection by Jupiter to bind the particle.

In order to find an upper limit to the energy from which particles may be captured, consider the most extreme encounter possible.  This is the case of a particle that has a tiny impact parameter with respect to Jupiter, and which has its initial velocity aligned with Jupiter's direction of motion.  Therefore, the particle's velocity with respect to Jupiter is
\begin{eqnarray}
	u = v - v_{\text{\jupiter}},
\end{eqnarray}
where $v = |\mathbf{v}|$.  The particle will be deflected through 180$^\circ$, so that 
\begin{eqnarray}
	u^\prime &=& -(v - v_{\text{\jupiter}}) \\
	v^\prime &=& 2 v_{\text{\jupiter}} - v.
\end{eqnarray}
The requirement that the particle is bound to the solar system after the scatter is equivalent to the statement
\begin{eqnarray}
	\left| v^\prime \right| \leq \sqrt{2}v_{\text{\jupiter}}.
\end{eqnarray}
Therefore, 
\begin{eqnarray}
	2v_{\text{\jupiter}} \leq v \leq (2 + \sqrt{2}) v_{\text{\jupiter}},
\end{eqnarray}
or
\begin{eqnarray}
	(2 - \sqrt{2}) v_{\text{\jupiter}} \leq v \leq 2v_{\text{\jupiter}},
\end{eqnarray}
and so 
\begin{eqnarray}
	E_{max} &\approx & \frac{1}{2} \left( 2 + \sqrt{2} \right)^2 v_{\text{\jupiter}}^2 - \frac{GM_\odot}{a_{\text{\jupiter}}} \\
	& = & 2\left( 1 + \sqrt{2} \right) v_{\text{\jupiter}}^2. \label{eq:emax}
\end{eqnarray}
This corresponds to a speed outside the gravitational sphere of influence of the Sun of
\begin{eqnarray}
	v_{s,max} &= & 2 \left( 1 + \sqrt{ 2 } \right)^{1/2} v_{\text{\jupiter}} \\
	    &= & 41 \text{ km/s}.
\end{eqnarray}
No WIMP with a speed far from the Sun that exceeds 41 km s$^{-1}$ with respect to the Sun can be gravitationally captured.

In addition to limiting the range of $E$ that we sample, we can also speed up the calculation by constraining the range of $J^2$ sampled.  This constraint is equivalent to specifying a range of orbital perihelia to probe, given that the perihelion $r_p$ is defined by  
\begin{eqnarray}
	E = \frac{1}{2} J^2 / r_p^2 - GM_\odot / r_p
\end{eqnarray}
such that the angular momentum for a given energy $E$ and perihelion is described by
\begin{eqnarray}
	J^2 (E, r_p) = 2 r_p \left( E r_p + GM_\odot \right).
\end{eqnarray}
The goal is to make the range of $r_p$ (and hence, $J^2$) large enough to encompass all orbits that might become bound to the solar system while keeping the range small enough so as not to waste computing resources by following unnecessary orbits.

We divide the gravitational scattering simulation into two parts, each defined by a different range of energy and perihelion: the ``Regular run'' and the ``High Perihelion run.''  The Regular run samples particle orbits with:
\begin{eqnarray}
	0 \leq E < v^2_\odot / 50 = \frac{1}{2}(44\text{ km/s})^2, &\mbox{ }r_p < 10 \text{ AU}.
\end{eqnarray}
The maximum perihelion of 10 AU was chosen to be large enough---twice the semi-major axis of Jupiter---so that this run would contain the vast majority of particles that are gravitationally captured.  If the Regular run misses any bound orbits due to the limit on $r_p$, those orbits should be found in the High Perihelion run, defined by
\begin{eqnarray}
	E < v^2_\odot / 50, &\mbox{ }10 < r_p < 20 \text{ AU}.
\end{eqnarray}

If we were to sample $E$ and $J^2$ according to the distribution of particle energy and angular momentum squared flowing in towards the Sun, Eq. (\ref{eq:gravflux}), the sampling probability would be:
\begin{eqnarray}
	G(E,J^2) \propto \label{eq:uniform}
		\begin{cases}
		f(\sqrt{2E}), \hbox{ }J^2 \in [J^2(E,r_p^{min}),J^2(E,r_p^{max})) \\
		0, \hbox{ }J^2 \in [J^2(E_{min}, r_p^{min}), J^2(E,r_p^{min})) \\
		 \text{ or } J^2 \in [J^2(E,r_p^{max}),J^2( E_{max}, r_p^{max})] 
		\end{cases} 	
\end{eqnarray}
in the range $E_{min} \leq E < E_{max}$ and $J^2(E_{min}, r_p^{min}) \leq J^2 < J^2( E_{max}, r_p^{max})$, where $r_p^{max}$ and $r_p^{min}$ are the maximum and minimum perihelia allowed in each run.  These ranges describe the maximum extent of $E$ and $J^2$ for any given run.  This sampling probability is highest in the high energy, high angular momentum part of the range considered.  However, we want to sample proportionally more low energy orbits in both the Regular and High Perihelion runs, since these are most easily captured.  We sample
\begin{eqnarray}
	G(E) = f(\sqrt{2E}), \label{eq:ge}
\end{eqnarray} 
in the range $E_{min} \leq E < E_{max}$, and uniformly sample $J^2(E, r_p^{min}) \leq J^2 < J^2( E, r_p^{max})$.  

We treat the halo WIMPs has having a Maxwellian DF in Galactocentric coordinates in the solar neighborhood,
\begin{eqnarray}
	f_h(\mathbf{x},\mathbf{v}_h,t) = \frac{n_\chi}{(2\pi \sigma^2)} e^{ - \mathbf{v}_h^2 / 2\sigma^2} \label{eq:local_maxwell},
\end{eqnarray}
where $v_h$ is the WIMP speed in Galactocentric coordinates, far outside the sphere of influence of the Sun.  $n_\chi = \rho_\chi / m_\chi$ is the local WIMP number density, where the $\rho_\chi \approx 0.3 \hbox{ GeV cm}^{-3}$ \cite{bergstrom1998b}.  We set the one-dimensional WIMP velocity dispersion $\sigma = v_\odot /\sqrt{2}$, where $v_\odot \approx 220 \hbox{ km s}^{-1}$ is the speed of the Local Standard of Rest \cite{gunn1979}.  In heliocentric coordinates, the DF is
\begin{eqnarray}
	f_s(\mathbf{x},\mathbf{v}_s,t) = f_h(\mathbf{x},\mathbf{v}_s + \mathbf{v}_\odot,t). \label{eq:local_maxwell_sun}
\end{eqnarray}
We use the angle-average of Eq. (\ref{eq:local_maxwell_sun}) to set the initial conditions. 

Once a sample particle's orbital parameters $E$ and $J^2$ are selected, its initial position is determined by randomly orienting the position vector to a point on a spherical shell with fixed radius $R$ relative to the Sun.  The initial speed vector is chosen to be oriented inward, with the angle $\theta$ relative to the position vector determined by $J^2$.  The speed $v$ is fixed by $R$ and $J^2$ since $J = Rv \sin \theta$.  The azimuth of the velocity vector relative to the position vector is also randomly chosen.  Thus, the initial position and velocity of the particle are completely determined.

\subsection{Coordinate System Choice}\label{sec:coords}

In general, we prefer to use heliocentric coordinates when integrating the equations of motion.  However, the symplectic integrator breaks down at large heliocentric distances.  This is because the gravitational potential in heliocentric coordinates has the form
\begin{eqnarray}
	\Phi(\mathbf{r},t) = -\frac{GM_\odot}{r} + \sum_i \Bigg[ -\frac{GM_i}{| \mathbf{r} - \mathbf{r}_i |} + \frac{GM_i \mathbf{r} \cdot \mathbf{r}_i}{r_i^3} \Bigg], \label{eq:helio_potential}
\end{eqnarray}
where $i$ denotes a planet.  At large heliocentric distances, the last term in Eq. (\ref{eq:helio_potential}), the indirect term, becomes large compared to the gravitational potential of the Sun.  Because we want to maintain $g(\mathbf{r},t) \approx | \mathbf{r} - \mathbf{r}_\odot |$ even at large distances from the Sun, we switch to barycentric coordinates far from the Sun.  We use a crossover radius $r_c = 53$ AU, which is point at which the value of the indirect term from Jupiter is approximately 10\% the gravitational potential of the Sun.  We find that using substantially smaller values of $r_c$ induces large energy errors due to frequent breaks to the symplectic integration scheme.

\subsection{Setting $h$}\label{sec:setting_h}

We initially set $h$ according to Table \ref{tab:grav_init}.  This is sufficient for a ``first pass'' through the solar system.  For long-term integrations, in order to both control errors near Jupiter and to speed up integration if particles settle onto tighter orbits, we reset $h$ after the particles pass through the Jupiter bubble.  $h$ is actually reset at the first aphelion after passing through the bubble since we have empirically determined that this is the point in the orbit at which a change of $h$ causes the minimum error.  We then set $h$ according to Table \ref{tab:gravh}.  Since the semi-major axis of an orbit can change substantially throughout the integration, it is useful to occasionally change $h$ to match $a$, either to speed up the integration or improve accuracy.  We allow $h$ to change after each passage through the Jupiter bubble up to 10 Myr; however, to control for errors caused by breaking the symplectic nature of the integrator repeatedly, we only allow $h$ to be reset if the energy changes by more than $20\%$ through a Jupiter bubble passage after that time.  Again, $h$ is always reset at aphelion.

\begin{table}
	\caption{The initial integration conditions for the gravitational capture simulation as a function of initial speed $v$ and Kepler perihelion $r_p$.  The values of $h$ are in units of $R^{-1}_\odot$ yr.}\label{tab:grav_init}
	\begin{center}
		\begin{ruledtabular}
		\begin{tabular}{r|r|r}
		Initial speed $v$ [km s$^{-1}$]	&$r_p < 5$ AU	&$r_p > 5$ AU \\
		\hline
		$v< 10$ 	&$2\times 10^{-7}$	&$3\times 10^{-7}$	\\
		$10 \le v < 20$	&$5\times 10^{-7}$	&$1\times 10^{-6}$	\\
		$ v \ge 20$	&$1 \times 10^{-6}$	&$2\times 10^{-6}$	\\ 
		\end{tabular}
		\end{ruledtabular}
	\end{center}
	
\end{table}

\begin{table}
	\caption{Choices for the fictitious time step $h$ as a function of semi-major axis for the gravitational capture simulations.  The semi-major axis refers to bound particles unless otherwise indicated.}\label{tab:gravh}
	\begin{center}
		\begin{ruledtabular}
		\begin{tabular}{c|c}
			$a$ range [AU] &$h[R^{-1}_\odot \text{ yr}]$ \\
			\hline
			$< 0.75$ &$10^{-4}$ \\
			$0.75 \le a < 1.1$ &$7\times 10^{-5}$ \\
			$1.1 \le a < 1.6$ &$6\times 10^{-5}$ \\
			$1.6 \le a < 3.5$ &$2\times 10^{-5}$ \\
			$3.5 \le a < 6.2$ &$1.5\times 10^{-5}$ \\
			$6.2 \le a < 13$ &$7\times 10^{-6}$ \\
			$13 \le a < 22$ &$10^{-6}$ \\
			$22 \le a < 30$ &$7\times 10^{-7}$ \\
			$30 \le a < 45$ &$6\times 10^{-7}$ \\
			$45 \le a < 120$ &$5\times 10^{-7}$ \\
			$120 \le a < 200$ &$4\times 10^{-7}$ \\
			$200 \le a < 500$ &$3\times 10^{-7}$ \\
			$a > 500$ or unbound &$2\times 10^{-7}$
		\end{tabular}
		\end{ruledtabular}
	\end{center}
	
\end{table}

\subsection{In the Sun}\label{sec:sun_treatment}
Once a WIMP is within $0.1 \hbox{ AU}$ of the Sun, we check if its perihelion will lie within $2R_\odot$ of the center of the Sun, where $R_\odot$ is the radius of the Sun.  If not, the symplectic integration continues without interruption.  If the WIMP does go through that region, though, we use the two-body map to evolve the WIMP through the region near the Sun.  If the WIMP goes through the Sun, we employ Monte Carlo techniques to determine if the WIMP scatters on a solar nucleus.  These techniques are further described in Appendix C of Paper I.

\subsection{Near Jupiter}\label{sec:jupiter_treatment}
We set the accuracy criterion to $| \Delta E / E | < 5 \times 10^{-7}$ at the point at which the WIMP exits the Jupiter bubble.  The bubble size was set to $l_{\jupiter} = 2.1\hbox{ AU}$ for particles with semi-major axes $ a < 100\hbox{ AU}$, and $l_{\jupiter} = 3.7$ AU for orbits with either $a > 100\hbox{ AU}$ or that were unbound.

\subsection{Stopping Conditions}\label{sec:stopping}
There were three circumstances in which orbit integrations were terminated: if the WIMP became unbound to the solar system, the WIMP rescattered onto an orbit of $a < 0.3$ (never to cross the Earth's path again), or the lifetime of the WIMP reached $t_\odot = 4.5$ Gyr the lifetime of the solar system.  

In Table \ref{tab:gravic}, we show how many WIMPs are simulated in each of the Regular and High Perihelion runs.  Simulations were performed using computational resources at Princeton University.  Each run required $\sim 10^5$ CPU-hours on dual-core $3.2$ GHz processors.

\begin{table}

	\caption{Gravitational scattering simulations}\label{tab:gravic}
	\begin{center}
	\begin{ruledtabular}
	\begin{tabular}{l|r}

 	Name &$N_p$ \\
	\hline

 	Regular &$4.8212 \times10^{9}$ \\
 	High Perihelion &$3.994 \times 10^{9}$ \\

 	\end{tabular}
	\end{ruledtabular}
	\end{center} 

\end{table}

\section{Distribution Functions}\label{sec:df}
We constructed the DFs using the procedure outlined in Appendix \ref{sec:df_estimator}.  For the Regular run simulation, the total (bound + unbound) DF was derived from a total of 369084 crossings within $z_{c} = 0.001$ AU of the Earth's orbit, a result of integrating $\approx 4.8\times 10^9$ particles with initial conditions distributed as in Eq. (\ref{eq:ge}).  Of those particles, 322441 particles were bound to the solar system for at least a short time, and 1224 of those bound orbits went through the Sun at least once.  However, not a single particle was elastically scattered in the Sun.  Of the 322441 particles that were bound to the solar system, only 5856 ever crossed the Earth's path ($R = 1$ AU, $|z| \le z_c$), of which 772 also went through the Sun.  Therefore, while only a small fraction of the bound orbits in this simulation contributed to the distribution function at the Earth, a large fraction of Sun-penetrating particles did.  

In the High Perihelion run (10 AU $< r_p < 20$ AU), there were only 9473 intersections with the Earth's orbit.  Of the nearly $4\times 10^9$ particle orbits simulated for the High Perihelion Run, 64559 became temporarily captured in the solar system, 335 contributed to the bound DF, and 54 went through the Sun.  As in the Regular run, none of the particles going through the Sun were scattered onto smaller orbits.  

A total of 70943 WIMP crossings from the two runs were used to build up the bound WIMP DF.

\begin{figure}
	\begin{center}
		\includegraphics[width=3.3in,angle=270]{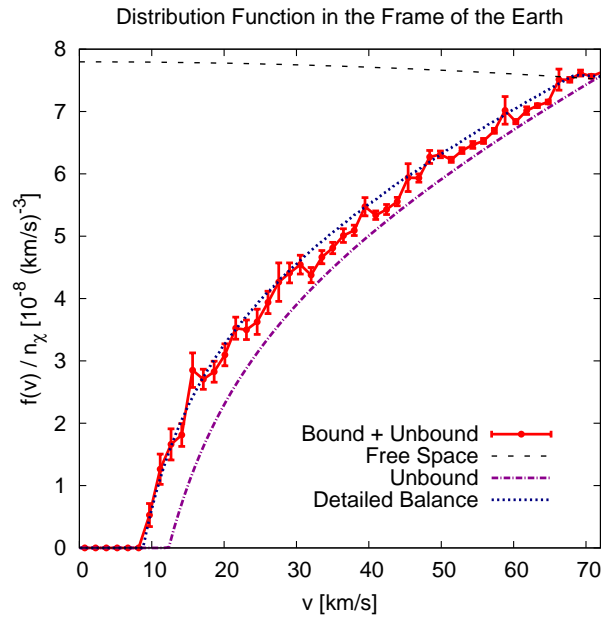}
	\end{center}
	\caption{The total distribution function from the gravitational capture simulations compared against several theoretical distribution functions.  }\label{fig:grav_theory}
\end{figure}

The WIMP DF from the simulations, including both bound and unbound WIMPs, is presented in Fig. \ref{fig:grav_theory}.  The DF is displayed in terms of the geocentric speed $v$ (the speed of WIMPs relative to the Earth in an inertial frame moving with the Earth) and divided through by the halo WIMP number density $n_\chi$.  It is normalized such that the number density of WIMPs near the Earth is given by $\int \hbox{d}v v^2 f(v)$.  We sum the DFs from each the Regular and High Perihelion runs and add the analytic DF of WIMPs (using Liouville's theorem) from the halo that were not in the energy and angular momentum windows used to set up the initial conditions for the simulations.  In this figure, we have also plotted the free space distribution function (the WIMP DF outside the sphere of influence of the Sun) and the DF of unbound WIMPs for comparison.

In this figure, we have also plotted the DF predicted by Gould's detailed balance argument, discussed in Section \ref{sec:intro_grav}.  This DF includes both unbound WIMPs and Jupiter-crossing bound WIMPs, assuming the latter have the same phase space density as the unbound WIMPs.  This is the most direct comparison to make with previous estimates of the DF; \citet{lundberg2004} estimate the WIMP DF in a solar system with more planets.  

The DF from the simulations is fairly well fit by the detailed balance DF at low speeds, but the fit is poor at higher speeds.  Moreover, we find that the DF of prograde bound WIMPs (those circulating in the same sense as Jupiter; the component of angular momentum perpendicular to the reference plane, $J_z$, is positive) is larger than the DF of retrograde WIMPs ($J_z < 0$).  Those DFs should be identical in the detailed balance approximation.  We believe these discrepancies are due to a violation of the key assumption in the detailed balance argument, that the phase space density of bound WIMPs depends on a single timescale, the angular diffusion timescale.

Instead, we find evidence that the timescale for ejection of WIMPs from the solar system is different from and shorter than the angular momentum diffusion timescale.  If only one timescale governed energy and angular momentum diffusion, we would expect that the distribution of the initial phase space coordinates of the WIMPs that built the bound DF at the Earth would be similar to the distribution of all bound WIMPs.  As an example, we would have expected the distribution of the initial WIMP angular momenta for all WIMPs bound to the solar system to look similar to the distribution of the initial WIMP angular momenta for WIMPs that contribute to the DF at the Earth.  In such a scenario, WIMPs with initially high angular momentum would have enough time to lose enough angular momentum so that the perihelia of the WIMPs would lie within the Earth's orbit for a time before being kicked out of the solar system.

\begin{figure}
	\begin{center}
		\includegraphics[width=3.3in]{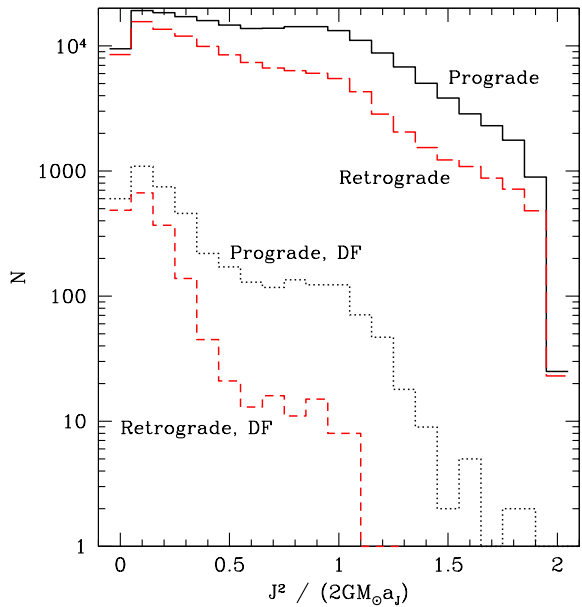}
	\end{center}
	\caption{\label{fig:angmom}The initial angular momentum distributions for all bound WIMPs (upper two curves), and the distribution of the initial angular momenta for bound WIMPs that contribute to the WIMP DF at the Earth (lower two curves).  The distributions are divided by whether the WIMPs were initially moving prograde or retrograde with respect to Jupiter.}
\end{figure}

We show these angular momentum distributions in Fig. \ref{fig:angmom}, separated by whether the WIMPs were initially prograde or retrograde.  In the figure, we show $\hat{J}^2 = J^2 / (2GM_\odot a_{\text{\jupiter}})$, the square of the WIMP specific angular momentum divided by the square of the angular momentum for a WIMP traveling at the escape speed and reaching perihelion at Jupiter's orbit.  We include all bound WIMPs in this plot, but we do not normalize the distribution to take into account the fact that we oversampled low energy orbits in the initial conditions (Section \ref{sec:start}).  If we had, the feature near $\hat{J}^2=1$ would be more prominent, as it corresponds with WIMPs with small impact parameters with respect to Jupiter.  Halo WIMPs with high initial energies must have small impact parameters on Jupiter in order to be captured to the solar system.  However, the qualitative differences between the bound WIMP and Earth-crossing bound WIMP initial angular momentum distributions are present in every energy interval.

We find that distribution of the initial WIMP angular momenta of the WIMPs in the DF at the Earth is skewed towards small angular momenta relative to the distribution of all bound WIMPs.  The high angular momentum WIMPs cannot lose enough angular momentum to reach the Earth's orbit before they are ejected from the solar system.  The effect is most pronounced for retrograde WIMPs.  

We considered that skew in the angular momentum distribution might be a result of the dependence of the WIMP lifetime in the solar system on the initial phase space coordinates.  The reason for believing this might be a significant effect is that the cross section for WIMP-planet encounters is a function of WIMP speed with respect to the planet.  Prograde WIMPs typically have small speeds with respect to the planet, and so the WIMP-planet cross section will typically be high.  Therefore, the timescale to eject a WIMP will be short.  For retrograde WIMPs, the relative speed of the WIMP increases as the angular momentum increases.  The WIMP-planet cross section should be small, and the lifetimes should be longer.  In general, WIMPs that are initially on prograde orbits will stay on prograde orbits, and likewise for retrograde orbits; Jupiter simply cannot move a WIMP with large, positive $J_z$ onto an orbit with large, negative $J_z$.  In the solar system as a whole, we find that the increase in the lifetime for retrograde WIMPs makes up for the smaller capture probability; there is an equal number of prograde and retrograde bound WIMPs in the solar system.

\begin{figure}
	\begin{center}
		\includegraphics[width=3.3in]{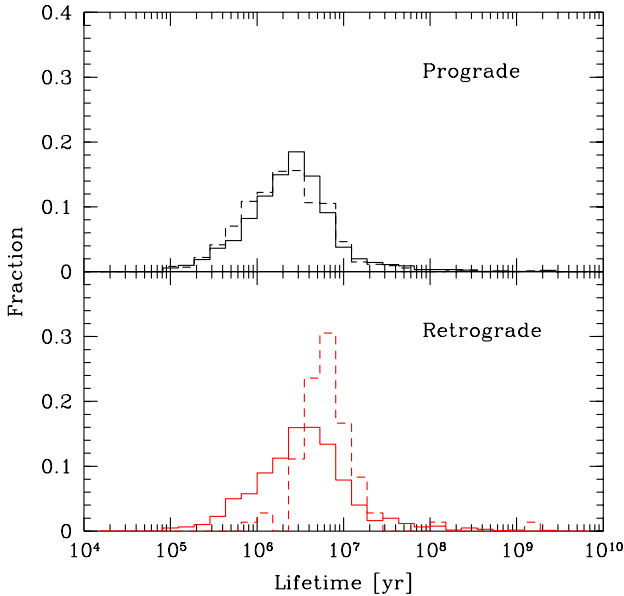}
	\end{center}
	\caption{\label{fig:angmom_lifetimes}Lifetime distributions for initially prograde or retrograde WIMPs.  The solid lines show the lifetime distributions for WIMPs with $\hat{J}^2 < 0.5$, and dashed curves mark those with $\hat{J}^2 > 0.5$.}
\end{figure}

However, the lifetime distribution of Earth-crossing WIMPs shows this not to be the case for the sample of Earth-crossing WIMPs.  In Fig. \ref{fig:angmom_lifetimes}, we plot the lifetime distributions of the WIMPs contributing the DF at the Earth as a function of the initial WIMP angular momentum.  We show the lifetime distributions for four WIMP populations: those initially prograde with low angular momentum (defined as $\hat{J}^2 < 0.5$ and $J_z > 0$); prograde with large angular momentum ($\hat{J}^2 > 0.5$, $J_z > 0$); retrograde with low angular momentum ($\hat{J}^2 < 0.5$, $J_z < 0$); and retrograde with high angular momentum ($\hat{J}^2 > 0.5$, $J_z < 0$).  We find that the median lifetimes for the WIMP populations are within a factor of three of each other, which is insufficient to explain the skewed angular momentum distributions in Fig. \ref{fig:angmom}.

Using these figures, we can also explain why there is a difference between the DF of prograde and retrograde orbits at the Earth.  In general, a WIMP that starts out on a prograde orbit stays prograde throughout its stay in the solar system, and a retrograde orbit stays retrograde.  The prograde WIMPs are much more successful than retrograde WIMPs at reaching sufficiently low angular momenta such that their perihelia may lie inside the Earth's orbit for some time, as demonstrated in Fig. \ref{fig:angmom}.  The prograde WIMP DF is larger than the retrograde WIMP DF because the angular momentum diffusion timescale is less for the prograde WIMPs, even though the typical ejection time for a retrograde WIMP is only slightly longer than for a prograde WIMP.

There is still the question of why the detailed balance DF fits the bound WIMP DF at small geocentric speeds.  To answer this question, we note that WIMPs populating this part of phase space tend to have small semi-major axes and have angular momentum vectors nearly aligned with Jupiter's but small enough in magnitude to intersect the Earth's orbit.  Using a crude diffusion argument, one can show that the timescale for large changes to the WIMP angular momentum should be similar to ejection timescale for these WIMPs, but not for other orbits.  Using an impulse approximation, the change to a WIMP's speed perpendicular to the direction of a planet in a planet-centric frame is
\begin{eqnarray}
	\delta u \sim \frac{GM_P}{bu},
\end{eqnarray}
where $M_P$ is the planet mass, $b$ is the impact parameter, and $u$ is the WIMP speed in the planet-centric frame.  If the WIMP orbit is nearly radial in a heliocentric frame, which we generally expect for Jupiter-crossing WIMPs that have perihelia inside the Earth's orbit, $u \approx \sqrt{ v^2 + v_P^2}$, where $v$ is the heliocentric WIMP speed and $v_P$ is the planet's circular speed about the Sun.  Thus, the change to the heliocentric WIMP speed is of order
\begin{eqnarray}
	\delta v \sim \frac{GM_P}{bv}.
\end{eqnarray}
The change to the WIMP energy is of order
\begin{eqnarray}
	\frac{\delta E}{E} = \frac{\delta a}{a } = \frac{a}{GM_\odot} v\delta v,
\end{eqnarray}
and the change to the angular momentum is
\begin{eqnarray}
	\delta J \sim a_P \delta v.
\end{eqnarray}

Using a random walk approximation, the change to either the energy or angular momentum (denoted as $X$ below) goes as
\begin{eqnarray}
	\langle (\Delta X)^2 \rangle \sim 10 N (\delta X)^2, \label{eq:delta_X}
\end{eqnarray}
where $N$ is the number of times a WIMP hits a planet in time $t$, and can be approximated by 
\begin{eqnarray}
	N \sim \frac{t}{(a_P/b)^2 P_\chi},
\end{eqnarray}
where $P_\chi$ is the WIMP orbital period.  The factor of $10$ in Eq. (\ref{eq:delta_X}) includes the Coulomb logarithm, which we have otherwise ignored in this simplified random walk calculation (for a more comprehensive treatment, see \cite{binney2008}).  

The rms change to the energy goes as
\begin{eqnarray}
	\langle (\delta E)^2 \rangle /E^2 \sim 10 \left( \frac{M_P}{M_\odot} \right)^2 \frac{t}{P_\chi},
\end{eqnarray}
so the ejection timescale goes as
\begin{eqnarray}
	t_{ej} \sim 0.1 \left( \frac{M_P}{M_\odot} \right)^{-2} P_\chi.
\end{eqnarray}
The rms change to the angular momentum goes as
\begin{eqnarray}
	\langle (\delta J) \rangle \sim 10 \left( \frac{GM_P}{v} \right)^2 \left(\frac{a_P}{a} \right)^2 \frac{t}{P_\chi},
\end{eqnarray}
and the timescale for a WIMP of $\hat{J} \sim 1$ to reach a completely radial orbit is
\begin{eqnarray}
	t_J \sim 0.1 \left( \frac{M_P}{M_\odot} \right)^{-2} \left( 2 - \frac{a_P}{a} \right) \left( \frac{ a}{a_P} \right)^2 P_\chi,
\end{eqnarray}
such that
\begin{eqnarray}
	\frac{t_J}{t_{ej}} \sim \left(2 - \frac{a_P}{a}\right) \left( \frac{a}{a_P}\right)^2.
\end{eqnarray}
Therefore, unless $a \sim a_P$, the timescale for large changes in the angular momentum $t_J$ will be much longer than the ejection timescale.  However, if $a \sim a_P$, the angular momentum diffusion timescale will be approximately the same as the ejection timescale.  Thus, the detailed balance assumption that the energy and angular momentum diffusion timescales are the same and equivalent to the angular diffusion timescale is met, and the WIMP DF should resemble the detailed balance DF at the lowest geocentric speeds.

In general, though, we find that while Jupiter is efficient at changing the energy of the WIMP orbits, it is quite inefficient at changing the WIMP perihelia.

A similar phenomenon arises in another context in the solar system.  It is thought that comets in the Oort Cloud originate in the outer solar system, $a=4-40$ AU.  Through interactions with the outer planets, the energy of the objects is pumped up, such that $ a \gtrsim 1000$ AU.  However, the perihelia stay nearly constant throughout this process.  External gravitational fields (from passing stars or molecular clouds) are required to move the perihelia of the comets outside the orbits of the planets \cite{heisler1986,duncan1987}.  Again, we see the discrepancy between the ejection timescale and the timescale to radically change the angular momentum of a body.

\subsection{Equilibrium Time}

In choosing to use the angle-averaged halo WIMP DF to set the initial conditions, we implicitly assumed that time equilibrium time for the bound WIMP DF (the time from the birth of the solar system beyond which the DF changes very little) was greater than the orbital period of the Sun about the Galactic Center, which is $\approx 200$ Myr.  The angle-averaged DF is approximately equal to the time-averaged DF due to the inclination of the plane of the solar system with respect to the Galactic plane (see Paper I and references therein).  If the equilibrium time is shorter than the Sun's orbital period about the Galactic Center, the procedure to determine the DF, as described in Appendix \ref{sec:df_estimator}, would need to incorporate a treatment of the anisotropy in the halo WIMP DF (Eq. \ref{eq:local_maxwell_sun}).

\begin{figure}
	\begin{center}
		\includegraphics[width=3.1in]{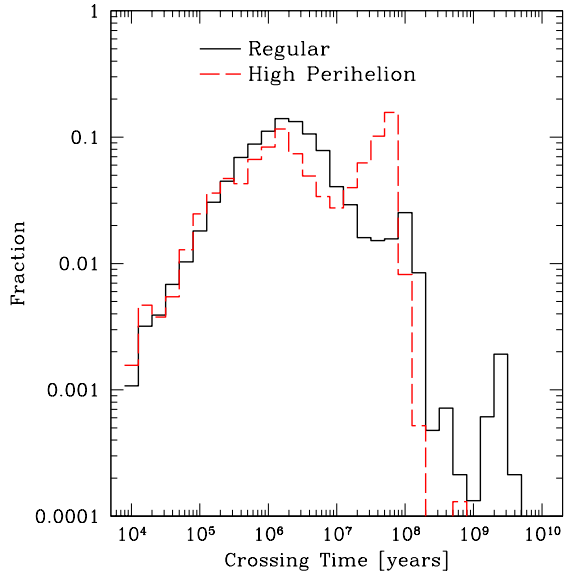}
	\end{center}
	\caption{(a) Distribution of times at which WIMPs cross the Earth's orbit.}\label{fig:gravlife}
\end{figure}

We show the distribution of times at which WIMPs cross the Earth's orbit ($R=1$ AU, $|z| < z_c$) in Fig. \ref{fig:gravlife}.  It is apparent that we have \emph{very few} crossing times past $\sim 200$ Myr, which would imply that the equilibrium timescale for the bound WIMP DF should be less than the orbital period of the Sun about the Galactic Center.  However, there is a possibility that the equilibrium time might be larger than it appears in Fig. \ref{fig:gravlife}.  While there is a sharp peak in the crossing time distribution near $\sim 1$ Myr, which is due to typical chaotic orbits, there some structure in the distribution of crossing times beyond $\sim 10$ Myr, and it is important to understand what types of orbits create this structure.  

Most of the Earth-orbit crossings beyond $\sim 10$ Myr are due to WIMPs that are temporarily stuck near mean-motion resonances, with a minority of the orbit-crossings coming from WIMPs initially captured onto large-$a$ orbits that scatter deep into the solar system at late times.  In addition, these WIMPs show signs of also being on Kozai cycles.  Kozai cycles are either librating or circulating solutions about a type of secular resonance in which the rate of perihelion precession $\dot{\omega}$ is small.  The characteristic behavior of such orbits includes large swings in eccentricity and inclination of an orbit while the semi-major axis remains roughly fixed.  The phenomenon of chaotic trajectories mimicking regular orbits near resonances for long times has been found in a number of systems, most of them two-dimensional \cite{karney1983,duncan1997,malyshkin1999}.  In the context of the solar system, such resonance-sticking has been found in simulations of Kuiper Belt objects (KBOs; \cite{duncan1997}) and comets \cite{malyshkin1999}.  The peak in the High Perihelion crossing time distribution near 50 Myr is due to a single resonance-sticking particle. 

Given that there are only five resonance-sticking WIMPs that contribute significantly to the WIMP DF at the Earth past 10 Myr, and that only one contributes past 100 Myr, it is clear that this long-lifetime tail in the DF is poorly sampled.  Since these few WIMPs account for $\sim 15\%$ of the bound WIMP Earth orbit crossings, it is important to understand how big (or small) the contribution of the resonance-sticking particles can be.  There are two important issues in estimating the possible size of the resonance-sticking DF: whether the orbits of the resonance-sticking WIMPs are typical of the population as a whole, and what the lifetime distribution of the WIMPs is.  The former can only be determined by more simulations.  There are perhaps some insights from previous work into the latter point.

The DF of resonance-sticking WIMPs beyond $\sim 100$ Myr can be estimated in the following way.  The rate at which a WIMP crosses the Earth's orbit can be described by $\dot{N}_c$, which should be constant as long as the WIMP is stuck to a resonance.  If the lifetime distribution is $N(>t) \propto t^{-\alpha}$, then the total DF beyond a time $t$ can be estimated by
\begin{eqnarray}
	f_{res}(>t) &\propto& \int_t N(t^\prime) \dot{N}_c(t^\prime) \hbox{d}t^\prime \\
		    &\propto &\begin{cases}  t^{1-\alpha}, & \alpha > 1 \\
					     \log(t_\odot/t), &\alpha = 1, \\
		    			     t_\odot^{1-\alpha}, &\alpha > 1.
		    \end{cases}
\end{eqnarray}
\citet{duncan1997}, in their simulations of KBOs in a solar system consisting of the four outer planets, and with orbits of KBOs restricted to the plane, find that $\alpha = 1$, in which case the resonance-sticking DF $f_{res}(>100\hbox{ Myr})$ should be a factor of several greater than what was found in our simulations.  Even in this case, the resonance-sticking WIMP DF will not be greater than the bound WIMP DF for typical chaotic WIMPs.  Previous work on planar systems, however, shows that usually $\alpha > 1$ \cite{karney1983,malyshkin1999}.  In that case, the only way we have underestimated the DF is if resonance-sticking WIMPs generically have a higher $\dot{N}_c$ than the WIMPs in our simulations.  We note that previous work on resonance-sticking has focused on two-dimensional systems, so any extrapolation to fully three-dimensional systems should be treated with caution.

For the gravitationally captured particles, the time averaging of the halo DF will turn out not to be justified if $\alpha > 1$.  If $\alpha \le 1$, the long-lifetime WIMPs will skew the equilibrium time higher, and so perhaps the time-averaging of the halo DF will be valid.  In the former case, while the results and interpretation here are qualitatively correct, to make a precise prediction of the distribution function of gravitationally bound particles in the solar system, one should use the original, anisotropic halo distribution function (Eq. \ref{eq:local_maxwell_sun}) to translate the WIMP initial conditions in the simulations to a DF.

\subsection{Loss Mechanisms}

There are two means by which particles may be lost to the solar system other than gravitational scatter by planets: 
\begin{itemize}
	\item Interactions with nuclei in the Sun (or, very rarely, the planets).  Even though no bound orbits were scattered in the Sun in any of the gravitational capture simulations, it is important to determine how the DF changes as a function of the strength of the dark matter-baryon interaction.  
	\item Interactions with external gravitational fields.  Galactic tides and encounters with distant stars become important for bound orbits with $a \gtrsim 1000$ AU.  Such Galactic gravitational fields are thought to be important in forming the Oort cloud as well as scattering Oort cloud comets into the solar system  \citep{heisler1986,duncan1987}.  It is important to understand how external fields will affect the distribution and lifetimes of WIMPs.
\end{itemize}

\begin{figure}
	\begin{center}
		\includegraphics[width=3.1in]{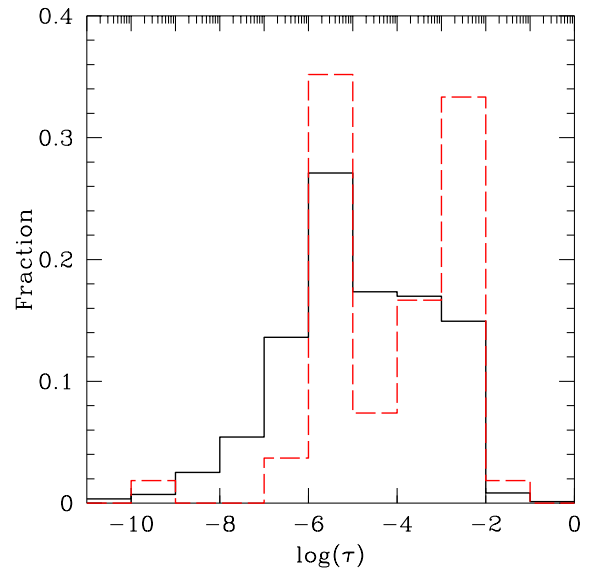}
	\end{center}
	\caption{The distribution of total optical depth per particle of particles that enter the Sun.  The solid line indicates the distribution for the Regular simulation, while the dashed line indicates that of the High Perihelion run.}\label{fig:od}
\end{figure}

In order to estimate the effect of the WIMP-nucleon cross section on the bound WIMP DF, we recorded the integrated optical depth as a function of time for each particle's orbit through the solar system.  Very few of the bound orbits (1224/322441) ever went through the Sun, but the optical depths $\tau$ of those that did are represented in Fig. \ref{fig:od}.  The median optical depth of Sun-crossing particles in the Regular run is $\tau_{med} \approx 10^{-5}$, and $\tau_{med} \approx 2\times 10^{-4}$ in the High Perihelion run.  Fig. \ref{fig:od} illustrates that very few particles have even a moderately high total optical depth if $m_\chi = 500$ AMU, $\sigma_p^{SI} = 10^{-43}$ cm$^2$, and $\sigma_p^{SD} = 0$. The WIMP-nucleon cross section (and hence, solar opacity) would need to be much, much higher in order for scattering in the Sun to rescatter any of the particles that pass through the Sun.  

\begin{figure}
	\begin{center}
		\includegraphics[width=3.3in,angle=270]{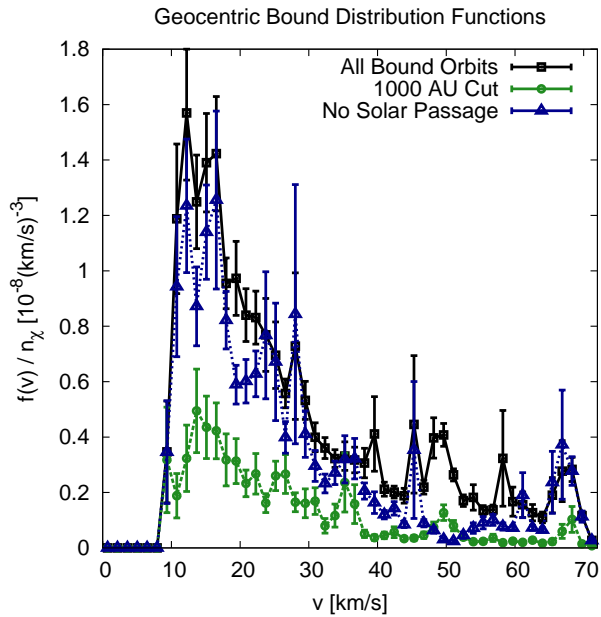}
	\end{center}
	\caption{The DF of bound particles from the gravitational capture simulations.  The squares mark the DF of all gravitationally bound WIMPs.  The circles indicate the DF of particles before they are lost to the solar system by Galactic tides.  The triangles indicate the DF of particles that never enter the Sun.}\label{fig:grav_loss}
\end{figure}

To determine the maximum effect of scattering in the Sun, we found the bound DF of only those particles that never enter the Sun.  This is represented by the triangles with error bars in Fig. \ref{fig:grav_loss}.  The majority of the bound WIMP DF is built up by particles that never enter the Sun.  Therefore, the DF of bound WIMPs at the Earth depends only weakly on the strength or type of the WIMP-baryon interaction.  

The effects of the external gravitational fields are independent of the WIMP mass and the WIMP-nucleon cross section.  In order to estimate the consequences of these these forces, we assumed that Galactic tides pull the perihelia of all orbits crossing outward through 1000 AU out of the solar system.  This is approximately the radius at which the timescale for external fields to remove orbital perihelia from the solar system is the same as the timescale for the planets to eject bodies \cite{duncan1987}.  In Fig. \ref{fig:grav_loss}, we show the DF arising from particle-Earth orbit intersections that occur before the particle passes outward through 1000 AU (circles).  The density of particles is noticeably lower that the total bound DF, generally by a factor of $\sim 3$.  It appears that WIMPs contributing to the DF at the Earth are initially captured on wide orbits that then shrink due to repeated encounters with Jupiter.  Even though our treatment of external gravitational fields is crude, the DFs in Fig. \ref{fig:grav_loss} indicate that torques from the Galactic tide should be included in estimates of the bound WIMP population at the Earth.

\subsection{Summary}\label{sec:df_summary}

The main results of these simulations are twofold.  First, the phase space density of unbound orbits is still quite a bit higher than that of the bound orbits above geocentric speeds $v \gtrsim 15$ km s$^{-1}$.  We expect that this will be true even if anisotropic initial conditions are used, the external Galactic gravitational potential is more accurately modeled, and once better statistics of bound orbits are obtained.  Second, the detailed balance DF is a poor fit to the WIMP DF for geocentric speeds $v \gtrsim 30\hbox{ km s}^{-1}$ due primarily to the difference in ejection and angular diffusion timescales, and secondarily to the presence of resonance-sticking orbits (points on the DF with larger-than-average error bars).  Third, the DF is largely insensitive to rescattering in the Sun.  Fourth, our crude treatment of Galactic gravitational fields suggests that these fields may be important in shaping the bound WIMP DF as well as the Oort cloud.

\begin{figure}
	\begin{center}
		\includegraphics[width=3.3in,angle=270]{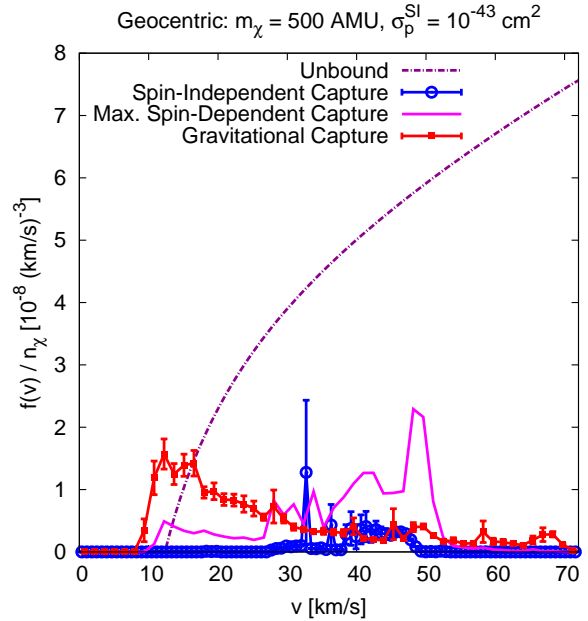}
	\end{center}
	\caption{Comparison of the geocentric bound distribution functions for $m_\chi = 500$ AMU and $\sigma_p^{SI} = 10^{-43}\hbox{ cm}^2$.  The dot-dashed line indicates the phase space density of unbound orbits.  The squares show the results from the gravitational capture simulations, the circles are the Large Mass DF ($\sigma_p^{SD} = 0$), and the solid magenta line indicates the estimated maximum DF resulting from spin-dependent scattering in the Sun ($\sigma_p^{SD} = 10^{-36}$ cm$^2$).}\label{fig:bound_comparison}
\end{figure}

Lastly, the phase space density of particles bound to the solar system by gravitationally scattering on Jupiter is generally higher than that of particles bound by elastic scattering in the Sun (``solar captured WIMPs'') for geocentric speeds $v < 30$ km s$^{-1}$ and $v > 50$ km s$^{-1}$.  This is demonstrated in Fig. \ref{fig:bound_comparison}, in which we show the results of the gravitational capture simulation as well as DFs from the solar capture simulations of Paper I.  If the spin-dependent cross section is high ($\sigma_p^{SD} \gtrsim 10^{-40} \hbox{ cm}^2$), then the bound DF for the elastically scattered particles may be higher than the gravitationally captured particles for $30 \text{ km s}^{-1} < v < 50$ km s$^{-1}$, especially if the gravitationally captured WIMP population is depleted by external forces.  However, the solar captured WIMP DF will be smaller or of approximately the same size as the gravitational capture DF if spin-independent scattering dominates in the Sun, or if $\sigma_p^{SD} \lesssim 10^{-40} \hbox{ cm}^2$.

\section{Direct Detection}\label{sec:dd}
Direct detection experiments look for nuclear recoil of rare WIMP-baryon interactions in the experimental target mass.  The WIMP-nucleus scattering rate per kg of detector mass per unit recoil energy $Q$ can be expressed as \citep[cf.][]{jungman1996}
\begin{eqnarray}
	\frac{\mathrm{d}R}{\mathrm{d}Q} = \left( \frac{m_A}{\mathrm{kg}} \right)^{-1}  \int_{v_{min}}^{\infty} \mathrm{d}^3 \mathbf{v} \frac{\mathrm{d} \sigma_A}{\mathrm{d} Q} v f(\mathrm{x},\mathrm{v}), \label{eq:drdq}
\end{eqnarray}
where $\mathrm{d}\sigma_A/\mathrm{d}Q$ is the differential interaction cross section between a WIMP and a nucleus of mass $m_A$ and atomic number $A$, and $\mathbf{v}$ is the velocity of the dark matter particle with respect to the experiment.   The lower limit to the integral in Eq. (\ref{eq:drdq}) is set to
\begin{eqnarray}
v_{min} = ( m_A Q / 2 \mu^2_A)^{1/2}, \label{eq:ch5vmin}
\end{eqnarray}
the minimum WIMP speed that can yield a nuclear recoil $Q$.  

We focus on direct detection rates for spin-independent interactions, but the results of this section can be applied qualitatively to spin-dependent interactions as well.  There is another class of direct detection experiment that is directionally sensitive \cite{alner2005b,naka2007,santos2007,nishimura2008,sciolla2008}.  In principle, the bound WIMPs should leave a unique signal in such experiments, but it would be challenging to distinguish this from the halo WIMPs given the small bound WIMP density, current errors in directional reconstruction, and high energy thresholds.

We find direct detection rates assuming $^{131}$Xe and $^{73}$Ge targets, since the current and planned experiments most sensitive to the spin-independent (and spin-dependent neutron) cross section have multiple isotopes of either xenon or germanium as their target mass.  We calculate the bound WIMP event rate for $m_\chi = 500$ AMU and $\sigma_p^{SI} = 10^{-43} \hbox{ cm}^2$.  The event rate can simply be scaled for lower (or higher) spin-independent cross sections.  The scaling for other values of $m_\chi$ and $\sigma_p^{SD}$ is different, but can easily be determined.

\begin{figure}
	\begin{center}
		\includegraphics[width=3.3in]{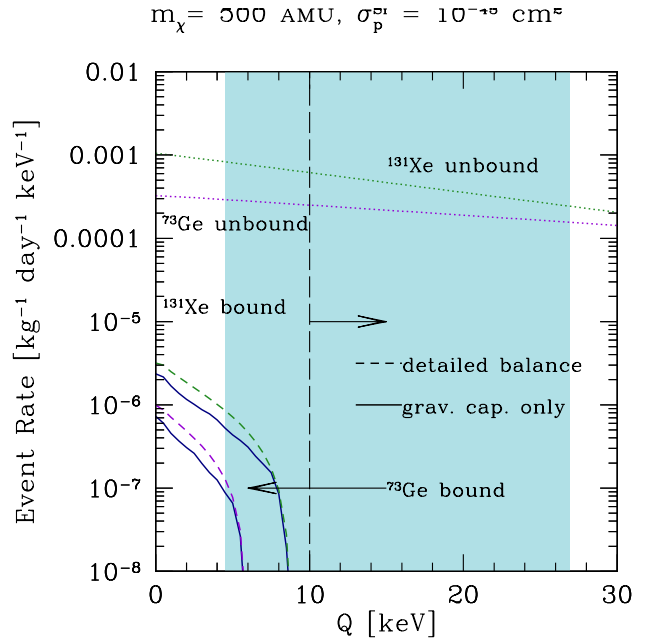}
	\end{center}
	\caption{\label{fig:dd_detailed}The differential direct detection signal from the halo, the gravitationally bound WIMP population, and the detailed balance estimate for the toy model solar system.  The shaded region indicates the XENON10 analysis region \cite{angle2008}, and the vertical dashed line indicates the lower limit to the CDMS analysis window (which extends to $Q=100$ keV) \cite{cdms2008}.}
\end{figure}

In Fig. \ref{fig:dd_detailed}, we show the differential direct detection event rate for the gravitationally bound WIMPs.  For comparison, we also show the event rate predicted for halo WIMPs using an angle-averaged Maxwellian speed distribution.  In addition, we show the direct detection rate for the detailed balance bound WIMP DF.  As expected, the event rate from the bound WIMPs in the simulation is somewhat less than predicted from detailed balance estimates.  The maximum contribution of the bound WIMPs to the event rate is at $Q = 0$, at which point it is approximately $\sim 0.3\%$ that of the halo.  In order to estimate the contribution of gravitationally bound WIMPs to the event rate in current experiments, we show the analysis windows for the XENON10 (shaded region) and CDMS (right of the vertical dashed line) experiments, which have xenon and germanium targets, respectively.  The CDMS experiment should be completely unaffected by bound WIMPs; $v_{min}$ is larger than the maximum bound WIMP speed at the analysis threshold.  The XENON10 experiment should be sensitive to bound WIMPs; however, the contribution to the total event rate is negligible.

\begin{figure}
	\begin{center}
		\includegraphics[width=3.3in]{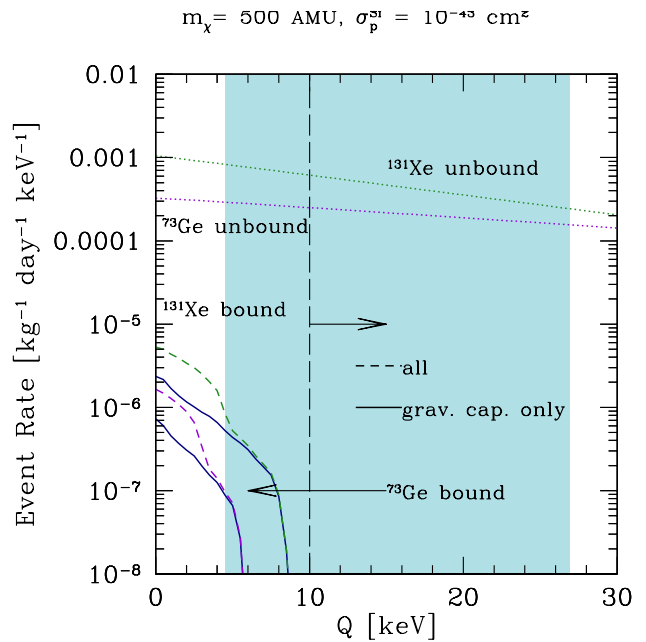}
	\end{center}
	\caption{The maximum contribution to the differential direct detection rate for $m_\chi = 500$ AMU and $\sigma_{p}^{SI} = 10^{-43}$ cm$^2$.  The upper lines represent the spin-independent event rate of halo WIMPs assuming a $^{131}$Xe and $^{73}$Ge) targets.  The lower solid lines show the event rate of gravitationally bound WIMPs; the lower dashed lines also include WIMPs bound to the solar system by solar capture assuming $\sigma_p^{SD} = 10^{-36}$ cm$^2$).}\label{fig:dd_max_grav}
\end{figure}

In Fig. \ref{fig:dd_max_grav}, we compare the direct detection rate of gravitationally bound WIMPs to the total bound WIMP event rate, which includes the maximum contribution to the event rate from the solar captured WIMPs discussed in Paper I.  The largest solar captured WIMP DF occurs for $m_\chi \sim$ a few hundred GeV, and if the spin-dependent WIMP-proton cross section $\sigma_P^{SD} \gtrsim 10^{-40} \hbox{ cm}^2$ and dominates the optical depth to WIMPs in the Sun.  In that case, we find that the bound WIMP event rate is dominated by solar captured WIMPs; the maximum value of the differential event rate also occurs at $Q = 0$, and is $\sim 0.5\%$ the halo event rate.  Since the solar captured DF is large at $30 < v < 50\hbox{ km s}^{-1}$ and small for other geocentric speeds, the gravitationally captured WIMPs dominate for $Q \gtrsim 5$ keV, which means they will dominate the bound WIMP signal in the XENON10 analysis window.

However, the solar captured WIMP DF is typically smaller unless $\sigma_p^{SD} \gtrsim 10^{-40}\hbox{ cm}^2$ or $\sigma_p^{SI} \gtrsim 10^{-42} \hbox{ cm}^2$.  For smaller WIMP-proton cross sections, gravitationally captured WIMPs will dominate the bound WIMP direct detection signal unless the Galactic tidal fields or the gravitational potentials are strong enough to severely reduce the gravitationally bound WIMP DF, as shown in Fig. \ref{fig:grav_loss}.

The main conclusion in this section is that the contribution of bound WIMPs to the event rate expected in direct detection experiments is negligible and will not affect parameter estimation based on the shape or normalization of the direct detection event rate.

\section{Neutrinos from WIMP Annihilation in the Earth}\label{sec:id}

WIMPs may accumulate and annihilate in the Earth.  The signature of WIMP annihilation will be GeV to TeV muon neutrinos, which may be observed in terrestrial neutrino observatories (e.g., Antares \cite{amram1999}, IceCube \cite{hill2006}) via the {\v C}erenkov radiation of muons created in charged-current interactions of muon neutrinos in and around the experiment.  In this section, we estimate the range of possible event rates due to WIMPs bound to the solar system assuming a neutralino WIMP.

The muon flux in the telescopes is proportional to the annihilation rate $\Gamma$ of WIMPs in the Earth.  If WIMPs quickly settle into an equilibrium distribution in the Earth once they are captured, the annihilation rate may be found by solving
\begin{eqnarray}
	\dot{N} = C - 2 \Gamma, \label{eq:dotN}
\end{eqnarray}
where $N$ is the number of WIMPs in the Earth. The capture rate of WIMPs in the Earth by elastic scattering is defined as
\begin{multline}
	C = \int \mathrm{d}^3 \mathbf{x} \int_{v_{f} < v_{esc}(\mathbf{x})} \mathrm{d}^3 \mathbf{v} \mathrm{d} \Omega  \sum_A \frac{\mathrm{d}\sigma_A}{\mathrm{d}\Omega} n_A (\mathbf{x})  v \\
	\times \, f(\mathbf{x},\mathbf{v},t). \label{eq:earthcap}
\end{multline}
Here, $\mathrm{d}\sigma_A / \mathrm{d}\Omega$ is the WIMP-nucleus elastic scattering cross section for nuclear species $A$ and $v$ is the relative speed between the WIMP and a nucleus.  The number density of species $A$ is described by $n_A(\mathbf{x})$.  The cutoff in the velocity integral reflects the fact that the WIMP's speed after scattering $v_f$ must be less than the local escape velocity $v_{esc}(\mathbf{x})$.  The second term in Eq. (\ref{eq:dotN}) is twice the annihilation rate 
\begin{eqnarray}
	\Gamma &=& \langle \sigma v \rangle_a \int \text{d}^3\mathbf{r}\, n^2(\mathbf{r},t)\\
		&= &\frac{1}{2} C_{a} N^2,
\end{eqnarray}
where $\langle \sigma v \rangle_a$ is the velocity-averaged annihilation cross section, $n(\mathbf{r},t) = N \tilde{n}(\mathbf{r},t)$ ($\int d^3\mathbf{r} \tilde{n} = 1$) is the density of WIMPs in the Earth.  The factor of two in Eq. (\ref{eq:dotN}) comes from the fact that most popular WIMP candidates are self-annihilating.  The coefficient $C_{a}$ is a constant as long as $\tilde{n}(\mathbf{r},t)$ is time-independent.  
\newline\indent
If the capture rate $C$ is time independent (i.e., the distribution function is time-independent), the annihilation rate can be calculated analytically:
\begin{eqnarray}
	\Gamma  = \frac{1}{2} C \tanh^2(t/t_{e}), \label{eq:gamma_analytic}
\end{eqnarray}
where 
\begin{eqnarray}
	t_{e} = ( C C_{a} )^{-1/2}
\end{eqnarray}
is the equilibrium timescale.  In the limit that the equilibrium timescale is small or large relative to the age of the solar system $t_\odot$, 
\begin{eqnarray}
	\Gamma = \begin{cases}
		\frac{1}{2} C &\text{if } t_\odot/t_{e} \gg 1 \\
		\frac{1}{2} C^2 C_{a} t^2_\odot &\text{if } t_\odot / t_{e} \ll 1. \label{eq:ann_nonequilibrium}
	\end{cases}
\end{eqnarray}

\begin{figure}
	\begin{center}
		\includegraphics[width=3.3in]{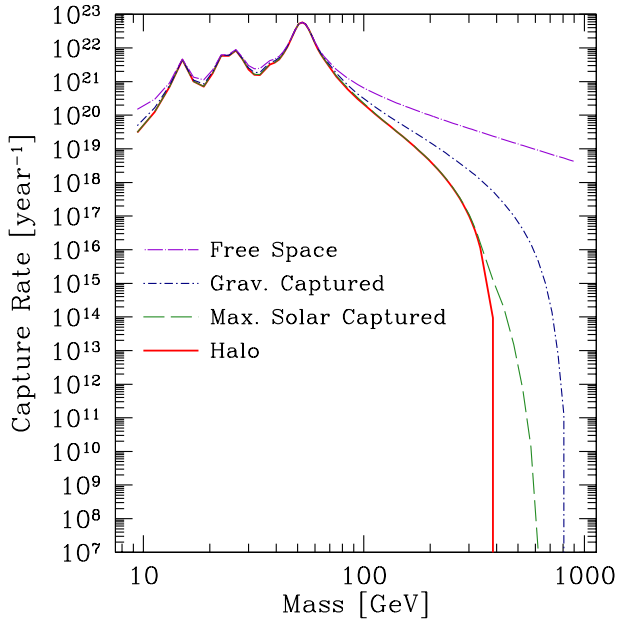}
	\end{center}
	\caption{Capture rate of WIMPs in the Earth as a function of WIMP mass for $\sigma^{SI}_{p} = 10^{-43}$ cm$^2$.}\label{fig:capearth}
\end{figure}

For parts of WIMP phase space not experimentally excluded, $t_e \gg t_\odot$, so the annihilation rate of WIMPs in the Earth is a sensitive function of the capture rate.  Since the Earth has a shallow potential well (the escape speed from the center of the Earth is $v_{esc} \approx 15 \hbox{ km s}^{-1}$), only low speed WIMPs may be captured by the Earth unless the WIMP mass is close to the mass of one of the dominant nuclear species in the Earth \cite{gould1988}.  Therefore, even though the bound WIMP population is small, it dominates the event rate in neutrino telescopes.  This is illustrated in Fig. \ref{fig:capearth}, in which we show capture rates assuming $\sigma_p^{SI} = 10^{-43}\hbox{ cm}^2$ using the Earth models in Refs. \citet{earth1,earth2}.  The capture rate from only unbound halo WIMPs in shown with the solid line, which drops to zero for $m_\chi > 400 $ GeV due to the cutoff in the DF at the escape speed from the solar system.  For reference, we have also plotted the capture rate for the free space DF, which the DF preferred by \citet{gould1991} assuming gravitational diffusion from all planets in the solar system.  This capture rate is substantially larger for $m_\chi \gtrsim 70$ GeV than that found in our simulations because of the relatively large free space WIMP phase space density at small geocentric speeds.  The maximum capture rate from solar captured WIMPs (with $\sigma_p^{SD}$ consistent with supersymmetry values) in addition to the unbound WIMPs in shown with the long-dashed line.  This capture rate is smaller than that for gravitationally captured WIMPs (dot-dashed line) because the gravitationally captured WIMP DF extends to lower speeds than the solar captured WIMP DF, and is generically larger for geocentric WIMP speeds $v < 30 \hbox{ km s}^{-1}$ (see Fig. \ref{fig:bound_comparison}).  The gravitationally captured WIMPs dominate the overall capture rate in the Earth for $m_\chi \gtrsim 100$ GeV.  We do not show the capture rate predicted by detailed balance because the DF is virtually indistinguishable from the simulation WIMP DF at the small speeds relevant for capture in the Earth.

To estimate a plausible range of muon event rates given the capture rates in Fig. \ref{fig:capearth}, we explore a hypersurface of minimal supersymmetric model (MSSM) parameter space assuming the WIMP is a neutralino.  We can in principle explore other models, but the MSSM yields, on average, somewhat larger spin-independent cross sections.  Given that iron is the most common element in the core of the Earth, and oxygen, silicon, and magnesium the most common element in the mantle, none of which has spin-dependent interactions with WIMPs, only in WIMP models with appreciable spin-independent interactions will capture in the Earth be relevant.  

To estimate the neutrino-induced muon event rate for neutrino telescopes from neutralino annihilation in the Earth, we use routines from the publicly available DarkSUSY v.5.0.2 code \citep{gondolo2004}.  Because searching the space of the large number of free parameters in the MSSM is a nearly impossible task, DarkSUSY has a simplified set of inputs from which all other MSSM parameters are set in a physically motivated way.  The seven free parameters, specified at the weak-breaking scale, are: $\mu$, the Higgsino mass parameter; $\tan \beta$, the ratio of the Higgs vacuum expectation values; $M_2$, the mass of one of the gauginos, through which the other two gaugino masses are specified; $m_{CP}$, the mass of the CP-odd Higgs (usually denoted by $m_\mathcal{A}$, which we avoid in order to prevent confusion with $m_A$, the mass of a nucleus with atomic number $A$); $m_0$, which sets the masses of the lepton and quark superpartners; and $A_t$ and $A_b$, which parametrize the strengths of the trilinear couplings in the most general MSSM Lagrangian.  

To generate a set of MSSM models for the neutralino, we scan a seven-dimensional hypersurface of the MSSM.  The range used for each parameter is given in Table \ref{tab:susy}.  For $\mu$, $M_2$, $m_{CP}$, $\tan \beta$, and $m_0$, we sample the range logarithmically, and sample the other parameters linearly in their ranges.  We accept a model if it makes it through the collider constraints, $0.05 < \Omega_\chi h^2 < 0.125$, and $\sigma_p^{SI} \ge 10^{-45}$ cm$^2$.  The upper limit on the allowed region of $\Omega_{\chi}h^2$ is approximately the $3\sigma$ range of $\Omega_{dm} h^2$ from the \emph{WMAP}-5 analysis \citep{komatsu2008}.  The lower limit is about half the $3\sigma$ lower limit from that analysis, since the neutralino may not be the only dark matter species.  We use 780 models which satisfied the requirements in our scans for the discussion below.

\begin{table*}
	\caption{Ranges of parameters for the seven-parameter DarkSUSY MSSM inputs at the weak scale}\label{tab:susy}
	\begin{center}
	\begin{ruledtabular}
	\begin{tabular}{lccccccc}
	
	SUSY parameters: &$\mu \text{ [GeV]}$	&$M_2\text{ [GeV]}$	&$m_{CP} \text{ [GeV]}$	&$\tan \beta$	&$m_0 \text{ [GeV]}$	&$A_t$	&$A_b$\\
	\hline
	min	&-50000	&-50000	&1	&1	&50	&-3	&-3\\
	\hline
	max	&50000	&50000	&50000	&60	&20000	&3	&3\\

	\end{tabular}
	\end{ruledtabular}
	\end{center}

\end{table*}

To estimate the muon event rate in a neutrino telescope, we set the muon energy threshold to $E^{th}_\mu = 1$ GeV.  This is somewhat optimistic for the IceCube experiment \citep{icecube2001,lundberg2004} unless muon trajectories lie near and exactly parallel to the PMT strings, but it is reasonable for the more densely packed water experiments (e.g., Super-Kamiokande).  We assume that the material surrounding the detector volume, the target material for neutrino interactions, is either water or ice, since the largest current and upcoming neutrino telescopes are immersed in oceans or the Antarctic ice cap.  We include all muons oriented within a 30$^\circ$ cone relative to the direction of the center of the Earth.  

\begin{figure*}
	\begin{center}
	\begin{tabular}{lll}
		(a) &(b) &(c) \\
		\includegraphics[width=2.3in]{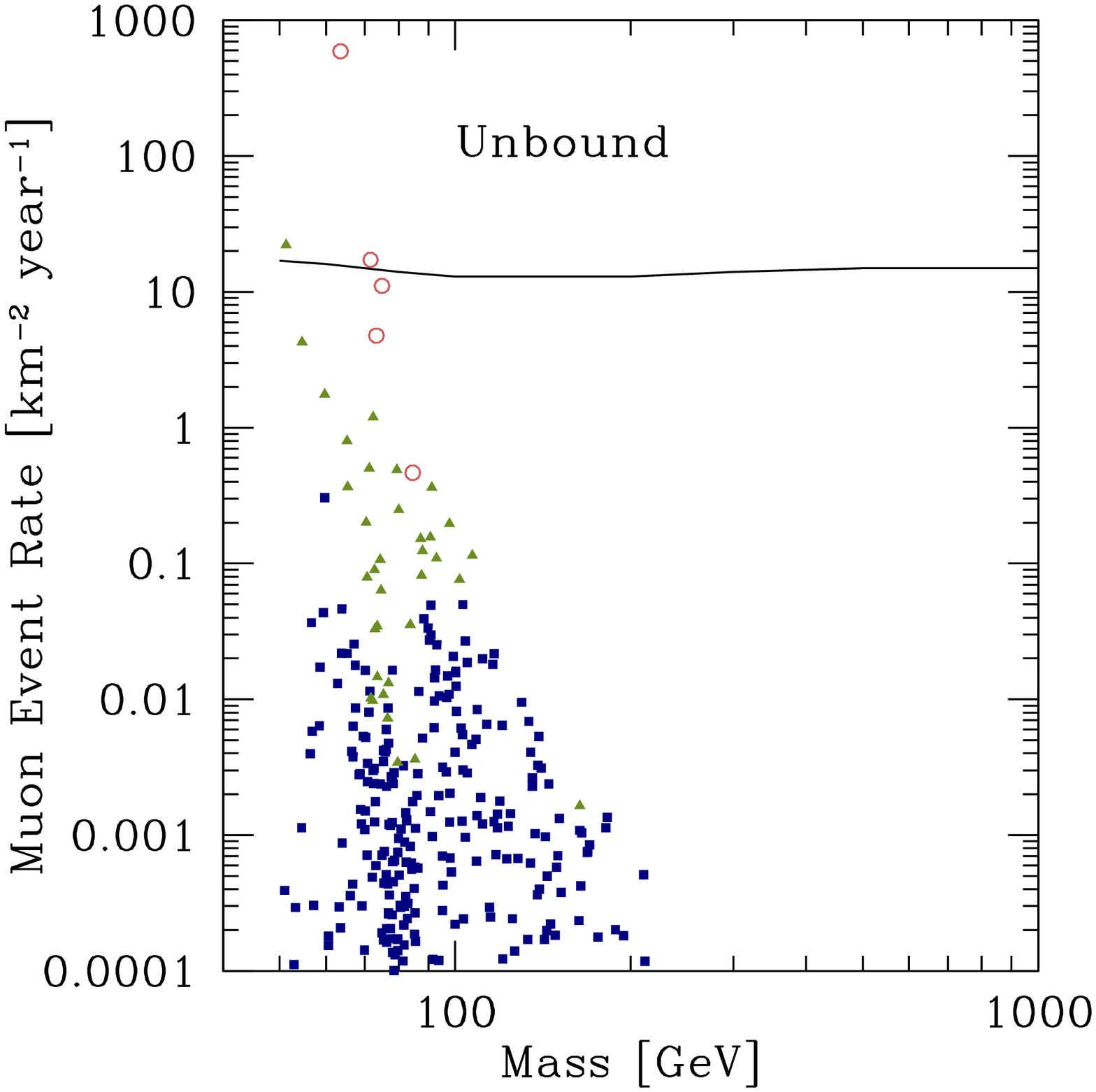} &\includegraphics[width=2.3in]{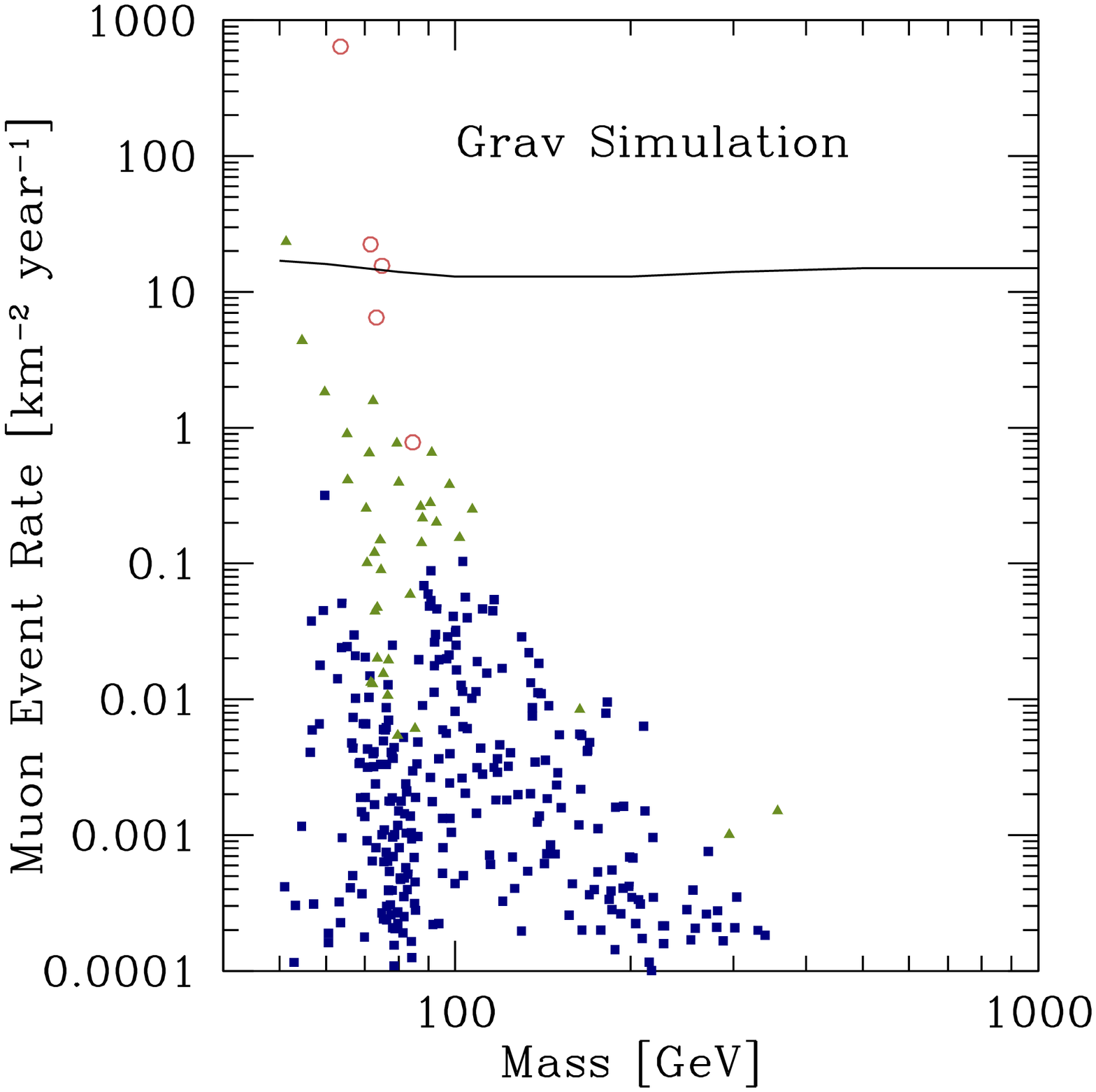} &\includegraphics[width=2.3in]{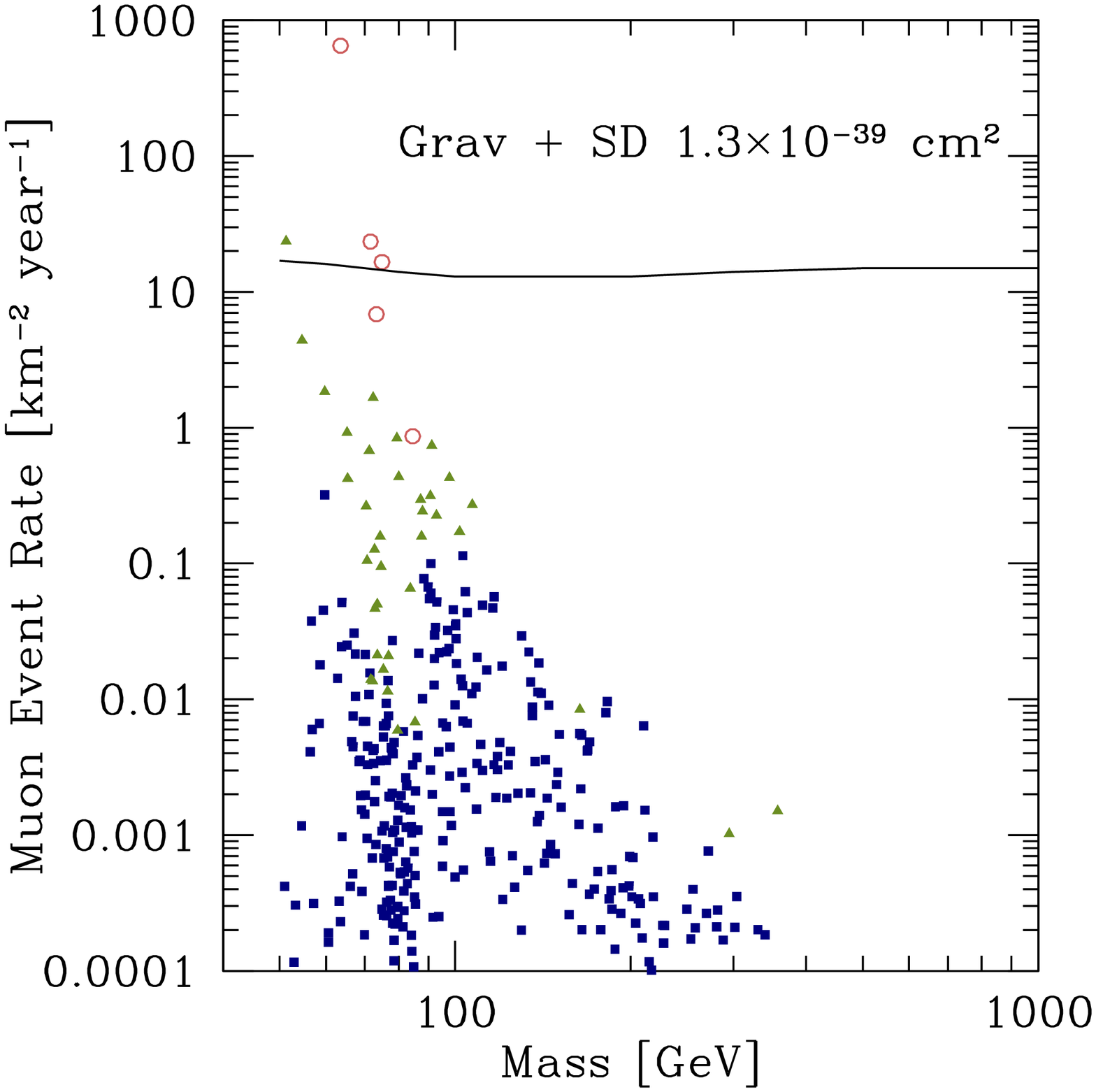} 
	\end{tabular}
	\end{center}
	\caption{\label{fig:neutrino_unbound}Muon event rates from (a) halo WIMPs unbound to the solar system, (b) halo and gravitationally bound WIMPs, and (c) halo and all bound WIMPs.  Open circles mark MSSM models for which $\sigma_p^{SI}$ is above the 2006 CDMS limit \cite{akerib2006}, filled triangles mark those with limits between that limit and the current best limits on $\sigma_p^{SI}$ (set by XENON10 for $m_\chi < 40$ GeV \cite{angle2008} and CDMS for $m_\chi > 40$ GeV \cite{cdms2008}), and filled squares denote models consistent with the best limits on elastic scattering cross sections.  The solid line is an optimistic detection threshold for the IceCube experiment \cite[][and references therein]{lundberg2004}.}
\end{figure*}

We present muon event rates in neutrino telescopes for various DFs in Fig. \ref{fig:neutrino_unbound}. We show the event rates for WIMPs unbound to the solar system, as well as for gravitationally captured WIMPs (in addition to the halo WIMPs) and for the maximum DF from solar capture (in addition to the halo and gravitationally captured WIMPs).  In the last case, we use the DF when $\sigma_p^{SD} = 1.3\times 10^{-39}\hbox{ cm}^2$, which is nearly the maximum spin-dependent cross section found in our parameter scans.  The solid black line in this figure represents the most optimistic flux threshold for IceCube \citep[][and references therein]{lundberg2004}.  To show how the event rates depend on the SUSY models for a given spin-independent cross section, we mark the models on the figure according to which direct detection experiments bracket the cross section for a given neutralino mass.  The open circles correspond to SUSY models with $\sigma_p^{SI}$ above the  that lie above the 2006 CDMS limit \cite{akerib2006}.  The triangles are models for which $\sigma_p^{SI}$ lies between the 2006 CDMS limit and the current best limits on $\sigma_p^{SI}$ (a combination of XENON10 \cite{angle2008} and CDMS \cite{cdms2008} limits), and squares denote models consistent with all current direct detection experiments.  

While we find that the bound WIMPs, especially those gravitationally captured to the solar system, do increase in the muon event rate in neutrino telescopes, the event rates fall far below threshold for the models in our scans.  While we cannot say that it is impossible for neutralino WIMPs to be observed by IceCube or other km$^3$-scale experiments, since we are only sampling a small part of the SUSY parameter space, the prospects do not look good.  If Galactic gravitational fields are important in the solar system for heliocentric distances as small as $\sim 1000$ AU, the event rate due to bound WIMPs would be a factor of $\sim 10$ smaller yet, since the DF of WIMPs before crossing outward through $r = 1000$ AU is a factor of $\sim 3$ smaller than in the absence of such fields for low geocentric speeds, and $\Gamma \propto C^2$ for such small capture rates.

As a caveat, only the flux of muons created outside the detector volume is calculated in DarkSUSY.  This was historically done because muon path lengths were long compared to the detector dimensions.  More recently, \citet{bergstrom1998} found that muons created inside the detector volume dominate the signal for smaller WIMP masses ($m_\chi \lesssim 300$ GeV) in large (km$^3$) telescopes, and that the expected event rate from muons created within the detector volume depends quite sensitively on the configuration of detectors inside that volume.  Therefore, the event rates used here ought to be considered a lower limit to the actual event rate in a large detector for neutralino masses $m_\chi \lesssim 300$ GeV.  From Fig. \ref{fig:neutrino_unbound}, we find that the event rate of muons created in the telescope volume would need to be at least two orders of magnitude larger than the event rate of muons created outside the telescope in order to be observed.  The ratio between the gravitationally bound WIMP event rate and the halo event rate, however, would be unchanged in this scenario.

\section{Discussion}\label{sec:discussion}

\subsection{Comparison with Previous Work}\label{sec:discussion_previous}
In this section, we discuss our results in comparison with previous work on gravitationally captured WIMPs, namely the work of \citet{gould1991} and \citet{lundberg2004}.  First, we focus on direct comparisons between our simulations and the work on the toy solar system.  Second, since the other authors considered the effects of planets, which we have not yet discussed in the context of the simulations, we describe what results from the toy solar system we expect will hold in the true solar system, and what we expect might change.

\emph{Toy solar system:}  First, we describe which of our results agree with previous work, and then how they disagree.  Our results agree with previous work in two important ways.  (i) While Lundberg \& Edsj{\"o} find that scattering in the Sun may drastically reduce the WIMP DF in a three-planet (Jupiter, Earth, Venus) solar system relative to the DF if the Sun were a point mass, they find that the WIMP DF in a one-planet (Jupiter) solar system is largely unaffected by the details of scattering in the Sun.  Our results (Fig. \ref{fig:grav_loss}) confirm their findings.  (ii) Even though our bound WIMP DF is smaller than the detailed balance DF, the two DFs are nearly identical for small geocentric WIMP speeds.  Since the capture rate of WIMPs in the Earth is most sensitive to the density of the lowest speed WIMPs, our predictions for the event rate for neutrino telescopes match those predicted using detailed balance arguments.

We find several important deviations from the detailed balance picture of Gould.  First, we showed that angular momentum and energy diffuse at different rates in the solar system.  This implied a deficit in the bound WIMP DF for large geocentric WIMP speeds, as well as an asymmetry between the prograde and retrograde WIMP DFs.  

Second, we found a set of long-lived ($t > 10$ Myr) resonance-sticking WIMPs that contributed $\sim 15\%$ of the bound WIMP DF.  Since our sample was small, we were unable to determine how statistically significant that contribution was.  The resonance-sticking bound WIMP DF depends on (i) the orbital properties of the WIMP while stuck to a resonance, (ii) the distribution of WIMPs among resonances, and (iii) the lifetime distribution.  The former two points will likely only be addressed in future, larger simulations.  We used the lifetime distributions from studies of comets in the solar system to argue that the resonance-sticking WIMP DF would be at most a similar size to the total gravitationally captured bound WIMP DF found in this work.  However, most of the work on comets was done in nearly planar systems, and with initial conditions for the comets that are quite different that the WIMP initial conditions \cite{duncan1997,malyshkin1999}.  We caution that WIMP orbits in the solar system are fully three-dimensional, and that the WIMP DF depends not on the overall lifetime distribution of WIMPs in the solar system, but on the lifetime distribution of WIMPs \emph{on Earth-crossing orbits}.  As an aside, all of the resonance-sticking WIMPs were originally captured onto orbits $a > 500$ AU, such that none of these WIMPs would have contributed to the WIMP DF if Galactic tidal fields were strong at $r \gtrsim 1000\hbox{ AU}$.  The only way to determine how important resonance-sticking is in the solar system is to perform larger simulations than those presented in this work, and incorporating a better treatment of external gravitational fields.

Finally, our work is the first to explore the possibility of Galactic gravitational fields affecting the bound WIMP DF.  Even though our treatment of the fields is crude (removing WIMPs from the inner solar system as soon as they pass outward through r = 1000 AU from the Sun), it suggests that external gravitational fields may play a significant role in shaping the bound WIMP distribution in the solar system.  Future simulations should include a more realistic treatment of the Galactic gravitational fields in order to make precision predictions for the bound WIMP DF at the Earth.

The potential importance of the Galactic gravitational fields brings up a shortcoming of treating all WIMP-planet encounters as local.  We find that the WIMP DF is much smaller when a crude treatment of Galactic tides is used because long-range capture of barely unbound WIMPs is quite important, and these WIMPs are typically captured onto barely bound orbits.  Therefore, the WIMP DF is sensitive to the details of capture in the solar system, which are not well-described by treating all encounters between WIMPs and planets as strictly local.

\emph{The effect of more planets on the DF:}  Our conclusions are based on simulations in a toy solar system, while the true solar system is far more complex.  The question is, how will putting Jupiter on a more realistic eccentric orbit and the presence of other planets alter the DF?  There are several things we expect.  First, we expect the asymmetry between prograde and retrograde orbits to persist in a more realistic solar system since all planets revolve in the same sense about the Sun as Jupiter.  Second, we expect the DF of resonance-sticking, Jupiter-crossing WIMPs of the types found in the toy solar system to be reduced, although it is not clear by how much.  The DF of such WIMPs is expected to be smaller because encounters with other planets (especially the outer planets) can perturb the WIMPs off the quasi-regular orbits they have when stuck to resonances.  Although simulations of KBOs have shown that up to 1\% of orbits can survive $\sim$ Gyr with a lifetime distribution $N(>t)\propto t^{-1}$ in a solar system consisting of only the four outer planets, we do not expect those results to generalize to WIMPs.  Even in the toy solar system, only $\sim 0.1\%$ of WIMPs contributing to the DF were on resonance-sticking orbits lasting $> 10$ Myr.  In addition, the typical WIMP perihelia are much smaller than the KBO perihelia, so that WIMPs may experience close encounters with gas giants other than Neptune (the planet which governs much of the KBO dynamics).

Finally, we expect the overall gravitationally bound WIMP number density to be within a factor of a few of what we found in the toy solar system, although this depends on whether a significant long lifetime tail (a result of various resonances involving WIMPs interior to Jupiter's orbit) exists.  The implications for direct detection experiments and neutrino telescopes will depend on how those WIMPs are distributed in velocity space.  Here, we consider a few possible ways in which a more complicated solar system might affect WIMP orbits and the size of the WIMP population.

First, we consider the capture of WIMPs to the solar system.  The gravitational cross section $\propto M_P^2$, where $M_P$ is the mass of planet $P$.  The inner planets are not nearly massive enough to significantly boost the capture rate.  In the outer solar system, Jupiter is the most massive planet by a factor of $\sim 3.3$, implying that its gravitational cross section is at least a factor of $\sim 10$ higher than any other planet.  Even if an outer planet were primarily responsible for capturing a WIMP to the solar system, there are two reasons why the WIMP DF at the Earth will be largely unaffected.  First, if an outer planet captures a WIMP on a Jupiter-crossing orbit, Jupiter will dominate the dynamics of the WIMP.  Second, Jupiter is the innermost outer planet.  Thus, the typical angular momentum of a WIMP captured by another planet will be higher than a WIMP captured by Jupiter.  We determined in Section \ref{sec:df} that the diffusion of angular momentum is typically much slower than energy diffusion.  Even if a WIMP with initially large angular momentum diffuses down to Jupiter's orbit, Jupiter will again dominate the dynamics of the WIMP, and will tend to eject the WIMP before it can diffuse further in angular momentum.  Therefore, even if an outer planet captures WIMPs to the solar system, it is unlikely to affect the DF of WIMPs at the Earth.

Next, we consider the combined effects of gravitational diffusion and rescattering in the Sun.  While Gould argued that the WIMP DF should be almost identical to the free space DF in the geocentric frame due to gravitational diffusion, \lu find that the DF is not much larger than the DF we found in our simulations.  In order to incorporate the effects of scattering in the Sun in the WIMP diffusion equation, \lu started 2000 WIMP orbits on a grid in geocentric velocity space, integrating orbits up to 49 Myr.  For each WIMP that hit in the Sun within that time, the original point in velocity space was assigned a scattering frequency ($\nu = 1/t_l$, where $t_l$ is the lifetime of the WIMP before scattering).  The loss term in the diffusion equation is $-n \nu$, where $n$ is the WIMP orbit density.  This means that WIMPs are lost from the solar system on the timescale for them to hit the Sun.  The treatment of scattering in \lu encompassed the various ways by which WIMPs are driven into the Sun.  Simulations of near-Earth object (NEO) and asteroid orbits suggest that secular resonances (occuring when either the rate of change of the longitude of perihelion $\dot{\varpi}$ or the longitude of the ascending node $\dot{\Omega}$ are the same as for one of the planets), as well as mean-motion and Kozai resonances, drive them into the Sun on $\sim 1-10$ Myr timescales \cite{farinella1994,gladman2000}.  

While these effects are incorporated into the diffusion equation of Lundberg \& Edsj{\"o}, we find that there are several ways in which a full orbit integration with a Monte Carlo treatment of scattering could improve on the work of Lundberg \& Edsj{\"o}, and reasons why the DF found by \lu is likely to be too small.  First, as we found in Paper I and in Section \ref{sec:df} of this work, WIMPs can survive many passages through the Sun before being removed from Earth-crossing orbits.  Therefore, the lifetimes of WIMPs could be substantially longer that assumed by Lundberg \& Edsj{\"o}.  The exact amount by which the lifetimes would be extended depends on how many passages WIMPs make through the Sun each time they are driven into the Sun by the planets, and how deeply into the Sun the WIMPs penetrate.  In addition, WIMP orbits may precess rapidly in the Sun, which may affect the type of orbit they are on.  This may change the frequency with which WIMPs are driven into the Sun in the future.

One effect that may decrease the DF is related to the short length of the Earth-crossing orbit simulations.  For a large swath in the geocentric velocity space, WIMPs were neither ejected from the solar system nor driven into the Sun on timescales less than $49$ Myr.  However, WIMPs may be driven into the Sun on longer timescales.  In particular, WIMPs on highly eccentric or highly inclined orbits can survive a long time before being scattered onto orbits that lead either to ejection or penetration of the Sun.  This is because the orbits are almost perpendicular to the direction of motion of the planets, leading to high speed encounters with the planets in which the WIMPs are not strongly deflected.  Therefore, \lu may be underestimating scattering in the Sun by not simulating such orbits long enough.  However, the size of the effect will be determined by the optical depth in the Sun to WIMPs, and what the typical integrated optical depth is each time a high eccentricity or high inclination WIMP is driven into the Sun.  If the integrated optical depth is small, then \lu may not be underestimating the effects of scattering for this population.  Underestimating the lifetime of the WIMPs that encountered the Sun within $49$ Myr will be a bigger effect than underestimating the scattering probability of the WIMPs that neither are ejected nor pushed into the Sun on 49 Myr timescales since the latter is closer to the lifetime of the solar system.

Finally, the distribution of times at which WIMPs enter the Sun for the first time is poorly sampled in \lu, and the system is chaotic, so interpolating the lifetimes among grid points in velocity space may not be the best way to interpolate lifetimes.  In Paper I, we found that the DF of solar captured WIMPs at the Earth was dominated by the long lifetime tail in the Earth-crossing WIMP distribution.  In fully integrating the WIMP orbits, the impact of the long lifetime, gravitationally captured WIMPs on the DF at the Earth can be understood.

In summary, we suspect that the DF predicted by \lu is too conservative; for most of WIMP parameter space consistent with experimental limits on the WIMP-baryon cross section, the optical depth per passage through the Sun is small.  Thus, WIMPs can survive many passages through the Sun before being removed from Earth-crossing orbits.  The effect of the longer lifetimes to scattering on the DF will also depend on the ejection timescale.  If the ejection timescale is shorter than the scattering timescale, the WIMP DF should resemble Gould's prediction of the DF if diffusion is the dominant gravitational process in the solar system, which would yield a bound WIMP number density not significantly larger than predicted in this work (although the fact that many of these WIMPs have low speeds has disproportionate consequences at neutrino telescopes).

However, diffusion is not necessarily the dominant gravitational process, at least in terms of the size of the DF.  Neither Gould nor \lu's treatment of the gravitational processes in the solar system incorporate resonances (\lu's treatment of resonances extends only to the timescale for WIMPs to enter the Sun; the treatment of WIMP-planet encounters is purely local).  While \citet{gladman2000} find for their sample of NEOs, resonances do not change the semi-major axis distribution relative to diffusion for $a < 2$ AU, they can affect the other orbital parameters.  Since the DF is sensitive to how WIMPs encounter the Earth, it is important to understand the differences between purely diffusive orbital evolution and the evolution when resonances are present.  Moreover, if the resonances can shield WIMPs from close encounters with planets (for example, the Kozai resonance assures that either the eccentricity or inclination of an orbit is high, making encounters with planets only occur at high speeds), they can extend the lifetimes of certain classes of orbits.  It is important to determine if resonances yield a significant long lifetime tail in the Earth-crossing WIMP distribution.

In conclusion, in a more realistic solar system, we expect that prograde WIMPs will have a higher density at the Earth than retrograde WIMPs, and that the density of resonance-sticking Jupiter-crossing WIMPs will be reduced.  We suspect that the DF of gravitationally bound WIMPs lies somewhere between the predictions of Gould and \lu in the absence of resonances, depending on the details of scattering in the Sun, unless Galactic gravitational fields affect the WIMP DF as strongly as suggested by our crude treatment in Section \ref{sec:df}.  Since resonances are important in the orbital evolution of NEOs and in the population of solar captured WIMPs described in Paper I, we suspect that they will also be important for gravitationally captured WIMPs whose orbits become interior to Jupiter's.  However, orbits will need to be integrated in the full solar system in order to evaulate the effects of these resonances on the distribution function.

\subsection{Future Simulations}
We would like to test these hypotheses with simulations of WIMPs in a more realistic solar system.  However, our experiences with simulating orbits in a toy solar system have highlighted some potential difficulties in going to a more complicated solar system.  As is the case for simulating the solar-captured WIMP population, the subject of Paper I, the primary problem will be to simulate a statistically significant number of orbits in finite computing time.  In the toy solar system, we required $\sim 2\times 10^5$ CPU-hours for the integration of $\sim 10^{10}$ WIMP orbits, of which only $\sim 3\times 10^5$ became bound, and only $\sim 6000$ of which contributed to the DF at the Earth.  There were only 5 resonance-sticking WIMPs with lifetimes longer than 10 Myr.  Since orbits in the real solar system display a much richer phenomenology than in the toy solar system, the number of WIMPs simulated in the toy solar system is insufficient to sample the spectrum of behavior in a more realistic solar system; in fact, we barely simulated enough WIMPs to find the resonance-sticking phenomenon in the toy solar system simulations.

Just as we did in Paper I (Section VIIB), we propose a few techniques for exploring solar system phenomenology in finite computing time.  First, as in Paper I, we recommend a series of intermediate simulations leading up to a full solar system simulation.  Perhaps the zero-order simulation in this series would include a more realistic treatment of Galactic gravitational fields in another toy solar system simulation, and a sufficiently large number of particles to determine the resonance-sticking WIMP DF.  The next simulation, in order to explore phenomena associated in systems with multiple planets, would include just three planets: Jupiter, Earth, and Venus.  The planets would initially be put on circular, coplanar orbits about the Sun to highlight diffusion phenomena as well as mean-motion and Kozai resonances.  Since the gravitational cross section scales as the square of the planet mass, we propose scaling up the masses of the Earth and Venus.  The masses of the planets could be reduced in follow-up simulations.  A possible next step would be to put the limited set of planets on more realistic orbits in order to explore changes to diffusion as well as the effects of secular resonances.

Second, we propose using more clever choices for the initial conditions of the simulations.  While we weighted the initial conditions towards low energy WIMPs and restricted the range of angular momenta, still only a small fraction of WIMPs were captured in the solar system in order to boost the yield of bound WIMPs.  We sampled WIMPs up to energies of $E \sim 0.5 (44 \hbox{ km s}^{-1})^2$.  Since the angular momenta were quite high in the High Perihelion run, the only WIMPs captured in the solar system had energies several orders of magnitude below this upper limit.  In the future, we recommend restricting High Perihelion-like simulations to lower energies.  In general, though, since WIMP orbits in the solar system are chaotic and the gravitational cross sections are small, it is difficult to further improve the yield of bound, Earth-crossing orbits by the choice of the weighting and range of $E$ and $J^2$ for the incoming halo WIMPs. 

However, if it appears that the long lifetime tail of the DF in the toy solar system and additional capture of WIMPs from the halo by the outer planets are unimportant, initial conditions for the simulations may be drawn from the toy solar system DF.  This is an attractive choice because the cross section for Jupiter to capture a halo WIMP is small (requiring us to simulate $\sim 10^{10}$ halo WIMPs to obtain a sample of $\sim 10^5$ bound WIMPs); by sampling the toy solar system DF, a far higher fraction of the simulated WIMPs should contribute to the full solar system WIMP DF at the Earth.  The reason to think this might work is because the typical diffusion timescales for the other planets in the solar system are far longer than for Jupiter.  Therefore, for a short period of time after the birth of the solar system, the toy solar system DF would be an accurate representation of the full solar system DF.  This technique is used in \citet{lundberg2004} to speed up their orbital diffusion calculations in a Jupiter-Venus-Earth solar system.  However, there are two potential drawbacks to this approach.  First, if the long lifetime tail in the toy solar system is important, the characteristic timescales of processes in the solar system related to planets other than Jupiter may approach the characteristic timescale in the DF, meaning that the long lifetime tail could be significantly altered by the other planets.  Second, if secular resonances act on short timescales to pump WIMPs into the Sun, as suggested in simulations of NEO orbits \cite{farinella1994,gladman2000}, the toy solar system DF will not be an accurate representation of the full solar system DF even on very short timescales.  Careful tests need to be performed to evaluate the possibility of this approach to the initial conditions.

In summary, we emphasize the need for a series of simulations, culminating in a full solar system simulation, in order to understand the importance of different phenomena in the solar system.  The choice of initial conditions is important to maximize statistics, but we are limited by the chaotic nature of the system.

\subsection{The Halo DF}\label{sec:discussion_df}
Gravitational capture is most effective for halo WIMPs with speeds $\sim 1\hbox{ km s}^{-1}$ outside the potential well of the Sun.  Therefore, it is important to determine the uncertainty on the halo DF at such small heliocentric speeds.  We briefly detail three ways in which deviations from our fiducial halo WIMP model may affect the gravitational capture rate $\dot{N}$ of WIMPs in the halo, and hence, the bound WIMP DF.  Since we discussed anisotropic halo speed distributions in Section \ref{sec:df}, we omit further reference to anisotropy here.

1.  \emph{Uncertainty in $v_\odot$, $\sigma$, and $\rho_\chi$:} The halo model (Eq. \ref{eq:local_maxwell}) used to set up the initial conditions (Section \ref{sec:start}) and to derive the WIMP DF from simulation outputs (Appendix \ref{sec:df_estimator}) is a single-variate Gaussian, non-rotating in a Galactocentric frame, and assumes that the speed of the Local Standard of Rest (LSR) $v_\odot = 220 \hbox{ km s}^{-1}$ and the one-dimensional halo WIMP velocity dispersion is $\sigma = v_\odot /\sqrt{2}$.  The uncertainty in the speed of the LSR is $\sim 10\%$ \cite{kerr1986}.  While the fiducial value of $v_\odot$ is what was used in our fiducial halo model, recent astrometry of masers in star-forming regions in the spiral arms of the Milky Way suggests $v_\odot \approx 250 \hbox{ km s}^{-1}$ \cite{reid2009}.  To determine what this uncertainty in $v_\odot$ implies for the uncertainty in the capture rate of WIMPs to the solar system, we find that the angle-averaged halo DF outside the sphere of influence of the Sun (average over Eq. \ref{eq:local_maxwell_sun}) 
\begin{eqnarray}
	f_s(v_s) \approx \frac{n_\chi}{(2\pi \sigma^2)^{3/2}} e^{-v^2_\odot/2\sigma^2}  
\end{eqnarray}
for the heliocentric speed $v_s \ll v_\odot$.  If $\sigma$ is fixed, a 10\% uncertainty in $v_\odot$ implies a $\sim 20\%$ uncertainty in the low speed halo WIMP DF, and hence, a 20\% uncertainty in the capture rate of WIMPs from the halo.  We note that we have neglected solar motion in our fiducial model, which is also of order 10\% of the speed of the Local Standard of Rest \cite{dehnen1998}.

However, we set $\sigma = v_\odot / \sqrt{2}$ in the halo DF.  Therefore, if we include the uncertainty of $\sigma$ through the uncertainty in $v_\odot$, we find that a 10\% uncertainty in $v_\odot$ yields a 30\% uncertainty in the DF.

There is further uncertainty in $\sigma$.  We used $\sigma = v_\odot / \sqrt{2}$ since this is the one-dimensional velocity dispersion for an isothermal halo DF for a power-law density distribution $n_\chi(r) \propto r^{-2}$.  Since halos are approximately described by an NFW profile,
\begin{eqnarray}
	n_\chi(r) \propto (r/r_s)^{-1} ( 1 + r/r_s)^{-2},
\end{eqnarray}
where $r_s$ is a scale radius, $n_\chi(r) \propto r^{-2}$ corresponds to the transition between the inner cusp ($n_\chi(r) \propto r^{-1}$) and the outer halo ($n_\chi(r) \propto r^{-3}$) \cite{navarro1997}.  Simulations suggest that the solar circle should be near this transition zone, although only small changes in the position of the Sun with respect to the transition radius can yield different relationships between the circular speed and the velocity dispersions (see Fig. 11 in \cite{moore2001}).  In general, a power-law density distribution $n_\chi(r) \propto r^{-\beta}$ yields a velocity dispersion $\sigma = v_\odot / \sqrt{\beta}$ for an isothermal DF, so if the Sun is well within the transition, $\sigma = v_\odot$, while $\sigma = v_\odot / \sqrt{3}$ if outside (see Appendix A of \cite{peter2008}).  

If there is no uncertainty in $v_\odot$ and it is fixed to the fiducial value, the uncertainty in the relation between $\sigma$ and $v_\odot$ yields an uncertainty in the low speed halo WIMP DF of order $40\%$.  Thus, the total uncertainty in the gravitational capture rate of WIMPs to the solar system due to uncertainties in $\sigma$ and $v_\odot$ is of order 50\% assuming that the WIMPs have an isotropic Maxwellian velocity distribution.

Simulations show that a multi-variate Gaussian is a good fit the macroscopic velocity distribution near the solar system at the speeds relevant for capture in the solar system, while we have used a single-variate Gaussian \cite{moore2001,helmi2002,vogelsberger2008}.  However, the velocity dispersion in each direction is within $\sim 10-20\%$ of the one-dimensional velocity dispersion used in our simulations.  Therefore, we believe that the uncertainty in $\beta$ yields greater uncertainty in the capture rate than treating the halo as a single-variate instead of a multi-variate Gaussian.

Finally, we consider uncertainties in the local WIMP density $\rho_\chi$.  Dark matter-only N-body simulations show that the density at any radius from the Galactic center should be within $\sim 10-20\%$ of its mean value in the shell.  However, the observed uncertainty in the local dark matter density is much higher because baryons dominate the local potential and density, and the distribution of baryons in the Galaxy is somewhat uncertain \cite{kuijken1989b,holmberg2000}.  There is approximately a factor of two uncertainty in the local dark matter density owing to the uncertainty in the baryon distribution \cite{bergstrom1998b}.  This yields a factor of $\sim 4$ uncertainty in the annihilation rate of WIMPs in the Earth.

2.  \emph{Halo substructure:} Finally, we consider the issue of dark matter substructure.  Substructure will have a small effect on the DF of gravitationally captured WIMPs for the same reasons why it has a negligible effect on the DF of solar captured WIMPs (discussed in Paper I).  Briefly, the probability that the solar system is in a dense subhalo is $\sim 10^{-4}$ at any given time \cite{vogelsberger2008}.  The equilibrium time of the gravitationally captured WIMP DF is sufficiently large that the solar system should have passed through many clumps in the equilibrium time.  The velocity distribution of subhalos is somewhat biased, at the level of the velocity dispersion of subhalos being $\sim 50\%$ higher than the smooth component \cite{diemand2004,faltenbacher2006}.  This yields order unity or less changes in the capture rate relative to the capture rate if the velocity distribution of subhalos were unbiased with respect to the smooth component.  Thus, the capture rate of WIMPs to the solar system averaged over the DF equilibrium time $\langle \dot{N} \rangle \approx \dot{N}$, where $\dot{N}$ is the capture rate of the smooth component in the absence of substructure.

Thus, we find that the presence of substructures in the halo and the uncertainty in $v_\odot$, $\sigma$, and $\rho_\chi$ of the smooth halo WIMP DF only change the overall capture rate of WIMPs in the solar system by at most of order unity.  This will yield a factor of a few uncertainty in the annihilation rate of WIMPs in the Earth, although it is unlikely that the rate will be high enough to produce an observable signal in IceCube. 

3.  \emph{Macroscopic dark matter structure:} In hydrodynamic N-body simulations of the evolution of disk galaxies, a second dark matter structure has been found.  This is a ``thick disk'' of dark matter, whose properties mimic the thick stellar disk, and is a result of satellite galaxies being preferentially dragged into the disk plane as they merge with disk galaxies \cite{lake1989,read2008,read2009}.  The thick disk has of order the same density at the solar circle as the more spherical dark matter halo, but is oblate and rotates in the same sense as the Sun about the Galactic Center.  The dark matter in the thick disk typically has a smaller speed with respect to the Sun and a smaller velocity dispersion than halo dark matter.  Both effects tend to boost the capture rate of WIMPs in the solar system.  The implications of this structure on the bound WIMP DF and event rates in neutrino telescopes are described in \citet{bruch2009}.

\section{Conclusion}\label{sec:conclusion}
The main conclusions of this work are:
\begin{itemize}

	\item We found the DF of WIMPs gravitationally bound to a toy model solar system consisting of Jupiter on a circular orbit about the Sun.

	\item While the DF matches the detailed balance prediction for geocentric speeds $v < 30\hbox{ km s}^{-1}$, the fit to that prediction is poor for larger speeds.  This discrepancy is largely due to the difficulty in populating retrograde, Earth-crossing orbits.  We find a small population of long-lived Earth-crossing resonance-sticking orbits, contributing $\sim 15\%$ to the number density of gravitationally bound WIMPs at the Earth.    

	\item The DF of gravitationally bound WIMPs is largely insensitive to the details of WIMP-nucleus interactions in the Sun, confirming the findings of \citet{lundberg2004}.

	\item If external Galactic gravitational fields become important for distances $r \gtrsim 1000$ AU from the Sun, the DF could be significantly smaller than shown in Fig. \ref{fig:bound_comparison}, by up to a factor of three.  Future simulations of WIMPs in the solar system should include more realistic treatments of external gravitational fields.

	\item The DF of gravitationally captured WIMPs is of order the same size as the largest DF of solar captured WIMPs consistent with limits on the WIMP-nucleon cross section (Paper I).  Since the DF of gravitationally captured WIMPs is larger at lower speeds, it dominates the annihilation rate of WIMPs in the Earth for $m_\chi \gtrsim 100$ GeV.  Since the DF of solar captured WIMPs is larger at higher geocentric speeds, solar captured WIMPs dominate the direct detection signal of bound WIMPs.

	\item The direct detection event rate of gravitationally captured WIMPs is never more than $\sim 0.1\%$ of the event rate of halo WIMPs.

	\item We find that the annihilation rate in the Earth of WIMPs gravitationally captured to the solar system agrees with \citet{gould1991}'s prediction for a toy solar system.  However, even though gravitationally captured WIMPs enhance the event rate of neutrinos from WIMP annihilation in the Earth's core, the signal falls far short of threshold for the IceCube experiment assuming a standard halo model.

\end{itemize}

\begin{acknowledgments}
We thank Scott Tremaine for advising this project, and Aldo Serenelli and Carlos Penya-Garay for providing tables of isotope abundances in the Sun. We acknowledge financial support from NASA grants NNG04GL47G and NNX08AH24G and from the Gordon and Betty Moore Foundation. The simulations were performed using computing resources at Princeton University supported by the Department of Astrophysical Sciences (NSF AST-0216105), the Department of Physics, and the TIGRESS High Performance Computing Center.  
\end{acknowledgments}

\appendix

\section{Estimating Distribution Functions}\label{sec:df_estimator}

In this Appendix, we describe how to estimate the WIMP DF at the Earth from the simulations, taking into account the weighting of initial conditions relative to the flux of halo WIMPs through the solar system.  Instead of estimating the WIMP DF at each point along the Earth's orbit, we find the DF averaged along the Earth's orbit, and within a height $z_c \ll a_\oplus$ of the reference plane (assuming the orbits of the Earth and Jupiter are coplanar).  By averaging over a small region along the Earth's orbit, we improve the DF statistics; by choosing $z_c$ to be small, the averaged DF should not be contaminated by gradients in the DF as a function of height above the reference plane.

To construct the averaged DF, we record the phase space coordinates each time a WIMP passes through a cylindrical shell of radius $a_\oplus$ and height $2z_c$ centered on the reference plane.  Thus, in effect, we record the unnormalized WIMP flux as a function of time.  To find the DF from these data, we must weight the flux with respect to the initial conditions of the simulation, and relate the flux to the DF.

The flux $F(\mathbf{v})$ is related to DF as follows.  The number of WIMPs passing outward through a wall of size $\delta A$ oriented in direction $\delta \hat{\mathbf{A}}$ in time $\delta t$ is
\begin{eqnarray}
	\delta N = \frac{\hbox{d}F}{\hbox{d}\mathbf{v}} \hbox{d} \mathbf{v} \delta A \delta t, \label{eq:delta_N_F}
\end{eqnarray}
where $v$ is the WIMP speed in a frame in which the wall is fixed.  Equivalently, the WIMPs that pass through the wall in time $\delta t$ inhabit a prism of base $\delta A$, long side $v\delta t$, and height $\delta t \mathbf{v} \cdot \delta \hat{ \mathbf{A}}$ with a number density $f(\mathbf{v})\hbox{d} \mathbf{v}$.  Therefore, in terms of the DF $f(\mathbf{v})$, the number of WIMPs passing outward through the wall is
\begin{eqnarray}
	\delta N &=& f(\mathbf{v}) \hbox{d} \mathbf{v} \delta A \delta t (\mathbf{v} \cdot \delta \hat{\mathbf{A}}) \\
		 &=& f(\mathbf{v}) v_\perp \hbox{d} \mathbf{v} \delta A \delta t, \label{eq:delta_N_f}
\end{eqnarray}
where $v_\perp$ is the component of WIMP speed perpendicular to the wall with outward normal $\delta\hat{\mathbf{A}}$.  Remembering that the DF is always positive, we equate Eqs. (\ref{eq:delta_N_F}) and (\ref{eq:delta_N_f}), we find

\begin{eqnarray}
	f(\mathbf{v}) = \left| \frac{\hbox{d}F(\mathbf{v})/\hbox{d}\mathbf{v}}{v_\perp } \right|, \label{eq:vperp}
\end{eqnarray}
where $v_\perp$ is component of the WIMP velocity perpendicular to the surface of the cylinder.

Therefore, if we find the differential flux $\hbox{d} F/\hbox{d} \mathbf{v}$, we can quickly determine the WIMP DF.  First, we must weight the WIMPs in the flux according to their initial conditions.  In Section \ref{sec:start}, we weighted the WIMP initial conditions relative to the flux of halo WIMPs through the solar system by a factor of
\begin{eqnarray}
	W(E) = \frac{ J^2 (E_{max},r_p^{max}) - J^2 (E_{min}, r_p^{min}) }{ J^2(E, r_p^{max}) - J^2(E, r_p^{min})}
\end{eqnarray}
assuming an isotropic halo WIMP velocity distribution.  Therefore, we weight the WIMPs in the flux by
\begin{eqnarray}
	w(E) = W^{-1}(E) = \frac{ J^2(E, r_p^{max}) - J^2(E, r_p^{min})}{ J^2 (E_{max},r_p^{max}) - J^2 (E_{min}, r_p^{min}) }. \label{eq:weight}.
\end{eqnarray}
This would be the proper weighting if the typical time for the WIMP DF to reach equilibrium were longer than the time for the Sun to circle the Galactic center, $\sim 200$ Myr, since the time averaged DF of Eq. (\ref{eq:local_maxwell_sun}) is nearly isotropic.  However, in Section \ref{sec:df}, the equilibrium timescale is less than this.  Thus, the particles should be weighted by their initial position and velocity relative to the orbital planet of the solar system and not just their initial speed (or energy).  For now, though, we treat the halo DF as isotropic.

The estimated average WIMP flux through the cylinder, properly normalized with respect to the initial particle distribution, is given by
\begin{multline}
	\frac{\hbox{d}\hat{F}(\mathbf{v},t)}{\hbox{d}\mathbf{v}} = \frac{\dot{N}}{2\pi a_\oplus z_c} \sum^{N_p}_{\alpha = 1} \sum^{N_\alpha}_{\beta = 1} w(E_\alpha) \delta^{(3)}( \mathbf{v} - \mathbf{v}_{\alpha\beta}) \Theta( t - t_{\alpha\beta}) \\
	/ \, \sum^{N_p}_{\alpha = 1} w(E_\alpha). \label{eq:flux_estimator}
\end{multline}
Here, $\dot{N}$ is the rate at which halo WIMPs in the energy and perihelion range considered in a simulation enter the solar system, $N_p$ is the total number of WIMP simulated in each run, $\alpha$ labels a particular WIMP, and $\beta$ denotes a particular passage of WIMP $\alpha$ through the cylinder. $\mathbf{v}_{\alpha\beta}$ is the velocity of the WIMP $\alpha$ during passage $\beta$, and $t_{\alpha\beta}$ is the time of that passage since the birth of the solar system.  The denominator in Eq. (\ref{eq:flux_estimator}) normalizes the flux.  To find the estimated bound WIMP DF, we insert Eq. (\ref{eq:flux_estimator}) into Eq. (\ref{eq:vperp}).

To find the DF, we set $z_c = 10^{-3} a_\oplus \approx 23.5 R_\oplus$.  The DF does not change within confidence limits for smaller values of $z_c$.  To estimate uncertainties the DF, we employ bootstrap resampling, drawing $N_p$ WIMPs with replacement from the initial WIMP sample.  For each simulation run, we do this 500 times.  To find the total DF of WIMPs bound to the solar system by gravitational capture, we sum the DFs from the Regular and High Perihelion runs and add the errors in quadrature, since the simulation runs are independent of each other.


\begin{thebibliography}{73}
\expandafter\ifx\csname natexlab\endcsname\relax\def\natexlab#1{#1}\fi
\expandafter\ifx\csname bibnamefont\endcsname\relax
  \def\bibnamefont#1{#1}\fi
\expandafter\ifx\csname bibfnamefont\endcsname\relax
  \def\bibfnamefont#1{#1}\fi
\expandafter\ifx\csname citenamefont\endcsname\relax
  \def\citenamefont#1{#1}\fi
\expandafter\ifx\csname url\endcsname\relax
  \def\url#1{\texttt{#1}}\fi
\expandafter\ifx\csname urlprefix\endcsname\relax\def\urlprefix{URL }\fi
\providecommand{\bibinfo}[2]{#2}
\providecommand{\eprint}[2][]{\url{#2}}

\bibitem[{\citenamefont{{Bertone} et~al.}(2005)\citenamefont{{Bertone},
  {Hooper}, and {Silk}}}]{bertone2005}
\bibinfo{author}{\bibfnamefont{G.}~\bibnamefont{{Bertone}}},
  \bibinfo{author}{\bibfnamefont{D.}~\bibnamefont{{Hooper}}}, \bibnamefont{and}
  \bibinfo{author}{\bibfnamefont{J.}~\bibnamefont{{Silk}}},
  \bibinfo{journal}{Phys. Rep.} \textbf{\bibinfo{volume}{405}},
  \bibinfo{pages}{279} (\bibinfo{year}{2005}), \eprint{arXiv:hep-ph/0404175}.

\bibitem[{\citenamefont{{Komatsu} et~al.}(2008)}]{komatsu2008}
\bibinfo{author}{\bibfnamefont{E.}~\bibnamefont{{Komatsu}}}
  \bibnamefont{et~al.} (\bibinfo{year}{2008}), \eprint{arXiv:0803.0547}.

\bibitem[{\citenamefont{{Jungman} et~al.}(1996)\citenamefont{{Jungman},
  {Kamionkowski}, and {Griest}}}]{jungman1996}
\bibinfo{author}{\bibfnamefont{G.}~\bibnamefont{{Jungman}}},
  \bibinfo{author}{\bibfnamefont{M.}~\bibnamefont{{Kamionkowski}}},
  \bibnamefont{and} \bibinfo{author}{\bibfnamefont{K.}~\bibnamefont{{Griest}}},
  \bibinfo{journal}{Phys. Rep.} \textbf{\bibinfo{volume}{267}},
  \bibinfo{pages}{195} (\bibinfo{year}{1996}), \eprint{arXiv:hep-ph/9506380}.

\bibitem[{\citenamefont{{Cheng} et~al.}(2002)\citenamefont{{Cheng}, {Feng}, and
  {Matchev}}}]{cheng2002}
\bibinfo{author}{\bibfnamefont{H.-C.} \bibnamefont{{Cheng}}},
  \bibinfo{author}{\bibfnamefont{J.~L.} \bibnamefont{{Feng}}},
  \bibnamefont{and} \bibinfo{author}{\bibfnamefont{K.~T.}
  \bibnamefont{{Matchev}}}, \bibinfo{journal}{Phys. Rev. Lett.}
  \textbf{\bibinfo{volume}{89}}, \bibinfo{pages}{211301}
  (\bibinfo{year}{2002}), \eprint{arXiv:hep-ph/0207125}.

\bibitem[{\citenamefont{{Hubisz} and {Meade}}(2005)}]{hubisz2005}
\bibinfo{author}{\bibfnamefont{J.}~\bibnamefont{{Hubisz}}} \bibnamefont{and}
  \bibinfo{author}{\bibfnamefont{P.}~\bibnamefont{{Meade}}},
  \bibinfo{journal}{\prd} \textbf{\bibinfo{volume}{71}},
  \bibinfo{pages}{035016} (\bibinfo{year}{2005}),
  \eprint{arXiv:hep-ph/0411264}.

\bibitem[{\citenamefont{{Arkani-Hamed}
  et~al.}(2006)\citenamefont{{Arkani-Hamed}, {Kane}, {Thaler}, and
  {Wang}}}]{arkanihamed2006}
\bibinfo{author}{\bibfnamefont{N.}~\bibnamefont{{Arkani-Hamed}}},
  \bibinfo{author}{\bibfnamefont{G.~L.} \bibnamefont{{Kane}}},
  \bibinfo{author}{\bibfnamefont{J.}~\bibnamefont{{Thaler}}}, \bibnamefont{and}
  \bibinfo{author}{\bibfnamefont{L.-T.} \bibnamefont{{Wang}}},
  \bibinfo{journal}{JHEP} \textbf{\bibinfo{volume}{8}}, \bibinfo{pages}{70}
  (\bibinfo{year}{2006}), \eprint{arXiv:hep-ph/0512190}.

\bibitem[{\citenamefont{{Baltz} et~al.}(2006)\citenamefont{{Baltz},
  {Battaglia}, {Peskin}, and {Wizansky}}}]{baltz2006}
\bibinfo{author}{\bibfnamefont{E.~A.} \bibnamefont{{Baltz}}},
  \bibinfo{author}{\bibfnamefont{M.}~\bibnamefont{{Battaglia}}},
  \bibinfo{author}{\bibfnamefont{M.~E.} \bibnamefont{{Peskin}}},
  \bibnamefont{and}
  \bibinfo{author}{\bibfnamefont{T.}~\bibnamefont{{Wizansky}}},
  \bibinfo{journal}{\prd} \textbf{\bibinfo{volume}{74}},
  \bibinfo{pages}{103521} (\bibinfo{year}{2006}),
  \eprint{arXiv:hep-ph/0602187}.

\bibitem[{\citenamefont{{Hooper} and {Baltz}}(2008)}]{hooper2008a}
\bibinfo{author}{\bibfnamefont{D.}~\bibnamefont{{Hooper}}} \bibnamefont{and}
  \bibinfo{author}{\bibfnamefont{E.~A.} \bibnamefont{{Baltz}}}
  (\bibinfo{year}{2008}), \eprint{arXiv:hep-ph/0802.0702}.

\bibitem[{\citenamefont{{Wai} et~al.}(2007)}]{wai2007}
\bibinfo{author}{\bibfnamefont{L.}~\bibnamefont{{Wai}}} \bibnamefont{et~al.},
  in \emph{\bibinfo{booktitle}{SUSY06}}, edited by
  \bibinfo{editor}{\bibfnamefont{J.~L.} \bibnamefont{{Feng}}}
  (\bibinfo{publisher}{American Institute of Physics, Melville, NY},
  \bibinfo{year}{2007}), vol. \bibinfo{volume}{903} of
  \emph{\bibinfo{series}{AIP Conference Series}}, pp.
  \bibinfo{pages}{599--602}.

\bibitem[{\citenamefont{{Adriani} et~al.}(2008)}]{pamela2008}
\bibinfo{author}{\bibfnamefont{O.}~\bibnamefont{{Adriani}}}
  \bibnamefont{et~al.} (\bibinfo{year}{2008}), \eprint{arXiv:0810.4995}.

\bibitem[{\citenamefont{{Chang} et~al.}(2008)}]{chang2008}
\bibinfo{author}{\bibfnamefont{J.}~\bibnamefont{{Chang}}} \bibnamefont{et~al.},
  \bibinfo{journal}{\nat} \textbf{\bibinfo{volume}{456}}, \bibinfo{pages}{362}
  (\bibinfo{year}{2008}).

\bibitem[{\citenamefont{{Aharonian} and {Neronov}}(2005)}]{aharonian2005}
\bibinfo{author}{\bibfnamefont{F.}~\bibnamefont{{Aharonian}}} \bibnamefont{and}
  \bibinfo{author}{\bibfnamefont{A.}~\bibnamefont{{Neronov}}},
  \bibinfo{journal}{Astrophys. Space Sci.} \textbf{\bibinfo{volume}{300}},
  \bibinfo{pages}{255} (\bibinfo{year}{2005}).

\bibitem[{\citenamefont{{Aharonian} et~al.}(2006)}]{aharonian2006}
\bibinfo{author}{\bibfnamefont{F.}~\bibnamefont{{Aharonian}}}
  \bibnamefont{et~al.}, \bibinfo{journal}{Phys. Rev. Lett.}
  \textbf{\bibinfo{volume}{97}}, \bibinfo{pages}{221102}
  (\bibinfo{year}{2006}), \eprint{arXiv:astro-ph/0610509}.

\bibitem[{\citenamefont{{Zaharijas} and {Hooper}}(2006)}]{zaharijas2006}
\bibinfo{author}{\bibfnamefont{G.}~\bibnamefont{{Zaharijas}}} \bibnamefont{and}
  \bibinfo{author}{\bibfnamefont{D.}~\bibnamefont{{Hooper}}},
  \bibinfo{journal}{Phys. Rev. D} \textbf{\bibinfo{volume}{73}},
  \bibinfo{pages}{103501} (\bibinfo{year}{2006}),
  \eprint{arXiv:astro-ph/0603540}.

\bibitem[{\citenamefont{{Profumo}}(2008)}]{profumo2008}
\bibinfo{author}{\bibfnamefont{S.}~\bibnamefont{{Profumo}}}
  (\bibinfo{year}{2008}), \eprint{arXiv:0812.4457}.

\bibitem[{\citenamefont{{Angle} et~al.}(2008{\natexlab{a}})}]{angle2008}
\bibinfo{author}{\bibfnamefont{J.}~\bibnamefont{{Angle}}} \bibnamefont{et~al.},
  \bibinfo{journal}{Phys. Rev. Lett.} \textbf{\bibinfo{volume}{100}},
  \bibinfo{pages}{021303} (\bibinfo{year}{2008}{\natexlab{a}}),
  \eprint{arXiv:astro-ph/0706.0039}.

\bibitem[{\citenamefont{{Angle} et~al.}(2008{\natexlab{b}})}]{angle2008b}
\bibinfo{author}{\bibfnamefont{J.}~\bibnamefont{{Angle}}} \bibnamefont{et~al.}
  (\bibinfo{year}{2008}{\natexlab{b}}), \eprint{arXiv:astro-ph/0805.2939}.

\bibitem[{\citenamefont{{CDMS Collaboration}}(2008)}]{cdms2008}
\bibinfo{author}{\bibnamefont{{CDMS Collaboration}}} (\bibinfo{year}{2008}),
  \eprint{arXiv:astro-ph/0802.3530}.

\bibitem[{\citenamefont{{Aprile} et~al.}(2002)}]{aprile2002}
\bibinfo{author}{\bibfnamefont{E.}~\bibnamefont{{Aprile}}} \bibnamefont{et~al.}
  (\bibinfo{year}{2002}), \bibinfo{note}{astro-ph/0207670},
  \eprint{arXiv:astro-ph/0207670}.

\bibitem[{\citenamefont{{Akerib} et~al.}(2006{\natexlab{a}})}]{akerib2006c}
\bibinfo{author}{\bibfnamefont{D.~S.} \bibnamefont{{Akerib}}}
  \bibnamefont{et~al.}, \bibinfo{journal}{Nucl. Instr. Meth. A}
  \textbf{\bibinfo{volume}{559}}, \bibinfo{pages}{411}
  (\bibinfo{year}{2006}{\natexlab{a}}).

\bibitem[{\citenamefont{{Hime}}(2007)}]{hime2007}
\bibinfo{author}{\bibfnamefont{A.}~\bibnamefont{{Hime}}}, \bibinfo{journal}{APS
  Meeting Abstracts} pp. \bibinfo{pages}{14002--+} (\bibinfo{year}{2007}).

\bibitem[{\citenamefont{{Gaitskell}}(2007)}]{gaitskell2007}
\bibinfo{author}{\bibfnamefont{R.}~\bibnamefont{{Gaitskell}}},
  \bibinfo{journal}{APS Meeting Abstracts} pp. \bibinfo{pages}{H3002+}
  (\bibinfo{year}{2007}), \bibinfo{note}{slides available at
  http://xenon.astro.columbia.edu/talks/\newline
  APS2007/070415\_DM\_Noble\_Gaitskell\_v08.pdf}.

\bibitem[{\citenamefont{{Desai} et~al.}(2004)}]{desai2004}
\bibinfo{author}{\bibfnamefont{S.}~\bibnamefont{{Desai}}} \bibnamefont{et~al.},
  \bibinfo{journal}{\prd} \textbf{\bibinfo{volume}{70}},
  \bibinfo{pages}{083523} (\bibinfo{year}{2004}),
  \eprint{arXiv:hep-ex/0404025}.

\bibitem[{\citenamefont{{Ackermann} et~al.}(2006)}]{ackermann2006}
\bibinfo{author}{\bibfnamefont{M.}~\bibnamefont{{Ackermann}}}
  \bibnamefont{et~al.}, \bibinfo{journal}{Astropart. Phys.}
  \textbf{\bibinfo{volume}{24}}, \bibinfo{pages}{459} (\bibinfo{year}{2006}).

\bibitem[{\citenamefont{{Bergstr{\"o}m}
  et~al.}(1998{\natexlab{a}})\citenamefont{{Bergstr{\"o}m}, {Ullio}, and
  {Buckley}}}]{bergstrom1998b}
\bibinfo{author}{\bibfnamefont{L.}~\bibnamefont{{Bergstr{\"o}m}}},
  \bibinfo{author}{\bibfnamefont{P.}~\bibnamefont{{Ullio}}}, \bibnamefont{and}
  \bibinfo{author}{\bibfnamefont{J.~H.} \bibnamefont{{Buckley}}},
  \bibinfo{journal}{Astropart. Phys.} \textbf{\bibinfo{volume}{9}},
  \bibinfo{pages}{137} (\bibinfo{year}{1998}{\natexlab{a}}),
  \eprint{arXiv:astro-ph/9712318}.

\bibitem[{\citenamefont{{Helmi} et~al.}(2002)\citenamefont{{Helmi}, {White},
  and {Springel}}}]{helmi2002}
\bibinfo{author}{\bibfnamefont{A.}~\bibnamefont{{Helmi}}},
  \bibinfo{author}{\bibfnamefont{S.~D.} \bibnamefont{{White}}},
  \bibnamefont{and}
  \bibinfo{author}{\bibfnamefont{V.}~\bibnamefont{{Springel}}},
  \bibinfo{journal}{\prd} \textbf{\bibinfo{volume}{66}},
  \bibinfo{pages}{063502} (\bibinfo{year}{2002}),
  \eprint{arXiv:astro-ph/0201289}.

\bibitem[{\citenamefont{{Read} et~al.}(2008)\citenamefont{{Read}, {Lake},
  {Agertz}, and {Debattista}}}]{read2008}
\bibinfo{author}{\bibfnamefont{J.~I.} \bibnamefont{{Read}}},
  \bibinfo{author}{\bibfnamefont{G.}~\bibnamefont{{Lake}}},
  \bibinfo{author}{\bibfnamefont{O.}~\bibnamefont{{Agertz}}}, \bibnamefont{and}
  \bibinfo{author}{\bibfnamefont{V.~P.} \bibnamefont{{Debattista}}},
  \bibinfo{journal}{Mon. Not. Roy. Astron. Soc.}
  \textbf{\bibinfo{volume}{389}}, \bibinfo{pages}{1041} (\bibinfo{year}{2008}),
  \eprint{arXiv:0803.2714}.

\bibitem[{\citenamefont{{Damour} and {Krauss}}(1999)}]{damour1999}
\bibinfo{author}{\bibfnamefont{T.}~\bibnamefont{{Damour}}} \bibnamefont{and}
  \bibinfo{author}{\bibfnamefont{L.~M.} \bibnamefont{{Krauss}}},
  \bibinfo{journal}{\prd} \textbf{\bibinfo{volume}{59}},
  \bibinfo{pages}{063509} (\bibinfo{year}{1999}),
  \eprint{arXiv:astro-ph/9807099}.

\bibitem[{\citenamefont{{Bergstr{\"o}m} et~al.}(1999)}]{bergstrom1999}
\bibinfo{author}{\bibfnamefont{L.}~\bibnamefont{{Bergstr{\"o}m}}}
  \bibnamefont{et~al.}, \bibinfo{journal}{JHEP} \textbf{\bibinfo{volume}{8}},
  \bibinfo{pages}{10} (\bibinfo{year}{1999}), \eprint{arXiv:hep-ph/9905446}.

\bibitem[{\citenamefont{{Gould} and {Alam}}(2001)}]{gould2001}
\bibinfo{author}{\bibfnamefont{A.}~\bibnamefont{{Gould}}} \bibnamefont{and}
  \bibinfo{author}{\bibfnamefont{S.~M.~K.} \bibnamefont{{Alam}}},
  \bibinfo{journal}{\apj} \textbf{\bibinfo{volume}{549}}, \bibinfo{pages}{72}
  (\bibinfo{year}{2001}), \eprint{arXiv:astro-ph/9911288}.

\bibitem[{\citenamefont{{Lundberg} and {Edsj{\"o}}}(2004)}]{lundberg2004}
\bibinfo{author}{\bibfnamefont{J.}~\bibnamefont{{Lundberg}}} \bibnamefont{and}
  \bibinfo{author}{\bibfnamefont{J.}~\bibnamefont{{Edsj{\"o}}}},
  \bibinfo{journal}{\prd} \textbf{\bibinfo{volume}{69}},
  \bibinfo{pages}{123505} (\bibinfo{year}{2004}),
  \eprint{arXiv:astro-ph/0401113}.

\bibitem[{\citenamefont{{Gould}}(1988)}]{gould1988}
\bibinfo{author}{\bibfnamefont{A.}~\bibnamefont{{Gould}}},
  \bibinfo{journal}{\apj} \textbf{\bibinfo{volume}{328}}, \bibinfo{pages}{919}
  (\bibinfo{year}{1988}).

\bibitem[{\citenamefont{{Gould}}(1991)}]{gould1991}
\bibinfo{author}{\bibfnamefont{A.}~\bibnamefont{{Gould}}},
  \bibinfo{journal}{\apj} \textbf{\bibinfo{volume}{368}}, \bibinfo{pages}{610}
  (\bibinfo{year}{1991}).

\bibitem[{\citenamefont{{Wisdom}}(1982)}]{wisdom1982}
\bibinfo{author}{\bibfnamefont{J.}~\bibnamefont{{Wisdom}}},
  \bibinfo{journal}{Astron. J.} \textbf{\bibinfo{volume}{87}},
  \bibinfo{pages}{577} (\bibinfo{year}{1982}).

\bibitem[{\citenamefont{{Farinella} et~al.}(1994)\citenamefont{{Farinella},
  {Froeschle}, {Gonczi}, {Hahn}, {Morbidelli}, and
  {Valsecchi}}}]{farinella1994}
\bibinfo{author}{\bibfnamefont{P.}~\bibnamefont{{Farinella}}},
  \bibinfo{author}{\bibfnamefont{C.}~\bibnamefont{{Froeschle}}},
  \bibinfo{author}{\bibfnamefont{R.}~\bibnamefont{{Gonczi}}},
  \bibinfo{author}{\bibfnamefont{G.}~\bibnamefont{{Hahn}}},
  \bibinfo{author}{\bibfnamefont{A.}~\bibnamefont{{Morbidelli}}},
  \bibnamefont{and} \bibinfo{author}{\bibfnamefont{G.~B.}
  \bibnamefont{{Valsecchi}}}, \bibinfo{journal}{\nat}
  \textbf{\bibinfo{volume}{371}}, \bibinfo{pages}{314} (\bibinfo{year}{1994}).

\bibitem[{\citenamefont{{Duncan} and {Levison}}(1997)}]{duncan1997}
\bibinfo{author}{\bibfnamefont{M.~J.} \bibnamefont{{Duncan}}} \bibnamefont{and}
  \bibinfo{author}{\bibfnamefont{H.~F.} \bibnamefont{{Levison}}},
  \bibinfo{journal}{Science} \textbf{\bibinfo{volume}{276}},
  \bibinfo{pages}{1670} (\bibinfo{year}{1997}).

\bibitem[{\citenamefont{{Gladman} et~al.}(2000)\citenamefont{{Gladman},
  {Michel}, and {Froeschl{\'e}}}}]{gladman2000}
\bibinfo{author}{\bibfnamefont{B.}~\bibnamefont{{Gladman}}},
  \bibinfo{author}{\bibfnamefont{P.}~\bibnamefont{{Michel}}}, \bibnamefont{and}
  \bibinfo{author}{\bibfnamefont{C.}~\bibnamefont{{Froeschl{\'e}}}},
  \bibinfo{journal}{Icarus} \textbf{\bibinfo{volume}{146}},
  \bibinfo{pages}{176} (\bibinfo{year}{2000}).

\bibitem[{\citenamefont{{Heisler} and {Tremaine}}(1986)}]{heisler1986}
\bibinfo{author}{\bibfnamefont{J.}~\bibnamefont{{Heisler}}} \bibnamefont{and}
  \bibinfo{author}{\bibfnamefont{S.}~\bibnamefont{{Tremaine}}},
  \bibinfo{journal}{Icarus} \textbf{\bibinfo{volume}{65}}, \bibinfo{pages}{13}
  (\bibinfo{year}{1986}).

\bibitem[{\citenamefont{{Duncan} et~al.}(1987)\citenamefont{{Duncan}, {Quinn},
  and {Tremaine}}}]{duncan1987}
\bibinfo{author}{\bibfnamefont{M.}~\bibnamefont{{Duncan}}},
  \bibinfo{author}{\bibfnamefont{T.}~\bibnamefont{{Quinn}}}, \bibnamefont{and}
  \bibinfo{author}{\bibfnamefont{S.}~\bibnamefont{{Tremaine}}},
  \bibinfo{journal}{Astron. J.} \textbf{\bibinfo{volume}{94}},
  \bibinfo{pages}{1330} (\bibinfo{year}{1987}).

\bibitem[{\citenamefont{{Mikkola} and {Tanikawa}}(1999)}]{mikkola1999}
\bibinfo{author}{\bibfnamefont{S.}~\bibnamefont{{Mikkola}}} \bibnamefont{and}
  \bibinfo{author}{\bibfnamefont{K.}~\bibnamefont{{Tanikawa}}},
  \bibinfo{journal}{Cel. Mech. Dyn. Astron.} \textbf{\bibinfo{volume}{74}},
  \bibinfo{pages}{287} (\bibinfo{year}{1999}).

\bibitem[{\citenamefont{{Preto} and {Tremaine}}(1999)}]{preto1999}
\bibinfo{author}{\bibfnamefont{M.}~\bibnamefont{{Preto}}} \bibnamefont{and}
  \bibinfo{author}{\bibfnamefont{S.}~\bibnamefont{{Tremaine}}},
  \bibinfo{journal}{Astron. J.} \textbf{\bibinfo{volume}{118}},
  \bibinfo{pages}{2532} (\bibinfo{year}{1999}),
  \eprint{arXiv:astro-ph/9906322}.

\bibitem[{\citenamefont{{Bahcall} et~al.}(2005)\citenamefont{{Bahcall},
  {Serenelli}, and {Basu}}}]{bahcall2005}
\bibinfo{author}{\bibfnamefont{J.~N.} \bibnamefont{{Bahcall}}},
  \bibinfo{author}{\bibfnamefont{A.~M.} \bibnamefont{{Serenelli}}},
  \bibnamefont{and} \bibinfo{author}{\bibfnamefont{S.}~\bibnamefont{{Basu}}},
  \bibinfo{journal}{Astrophys. J.} \textbf{\bibinfo{volume}{621}},
  \bibinfo{pages}{L85} (\bibinfo{year}{2005}), \eprint{arXiv:astro-ph/0412440}.

\bibitem[{\citenamefont{{Gunn} et~al.}(1979)\citenamefont{{Gunn}, {Knapp}, and
  {Tremaine}}}]{gunn1979}
\bibinfo{author}{\bibfnamefont{J.~E.} \bibnamefont{{Gunn}}},
  \bibinfo{author}{\bibfnamefont{G.~R.} \bibnamefont{{Knapp}}},
  \bibnamefont{and} \bibinfo{author}{\bibfnamefont{S.~D.}
  \bibnamefont{{Tremaine}}}, \bibinfo{journal}{Astron. J.}
  \textbf{\bibinfo{volume}{84}}, \bibinfo{pages}{1181} (\bibinfo{year}{1979}).

\bibitem[{\citenamefont{{Binney} and {Tremaine}}(2008)}]{binney2008}
\bibinfo{author}{\bibfnamefont{J.}~\bibnamefont{{Binney}}} \bibnamefont{and}
  \bibinfo{author}{\bibfnamefont{S.}~\bibnamefont{{Tremaine}}},
  \emph{\bibinfo{title}{{Galactic Dynamics}}} (\bibinfo{publisher}{Princeton,
  NJ, Princeton University Press}, \bibinfo{year}{2008}).

\bibitem[{\citenamefont{{Karney}}(1983)}]{karney1983}
\bibinfo{author}{\bibfnamefont{C.~F.~F.} \bibnamefont{{Karney}}},
  \bibinfo{journal}{Physica D Nonlinear Phenomena}
  \textbf{\bibinfo{volume}{8}}, \bibinfo{pages}{360} (\bibinfo{year}{1983}),
  \eprint{arXiv:nlin/0501023}.

\bibitem[{\citenamefont{{Malyshkin} and {Tremaine}}(1999)}]{malyshkin1999}
\bibinfo{author}{\bibfnamefont{L.}~\bibnamefont{{Malyshkin}}} \bibnamefont{and}
  \bibinfo{author}{\bibfnamefont{S.}~\bibnamefont{{Tremaine}}},
  \bibinfo{journal}{Icarus} \textbf{\bibinfo{volume}{141}},
  \bibinfo{pages}{341} (\bibinfo{year}{1999}), \eprint{arXiv:astro-ph/9808172}.

\bibitem[{\citenamefont{{Alner} et~al.}(2005)}]{alner2005b}
\bibinfo{author}{\bibfnamefont{G.~J.} \bibnamefont{{Alner}}}
  \bibnamefont{et~al.}, \bibinfo{journal}{Nucl. Instr. Meth. A}
  \textbf{\bibinfo{volume}{555}}, \bibinfo{pages}{173} (\bibinfo{year}{2005}).

\bibitem[{\citenamefont{{Naka} et~al.}(2007)}]{naka2007}
\bibinfo{author}{\bibfnamefont{T.}~\bibnamefont{{Naka}}} \bibnamefont{et~al.},
  \bibinfo{journal}{Nucl. Instr. Meth. A} \textbf{\bibinfo{volume}{581}},
  \bibinfo{pages}{761} (\bibinfo{year}{2007}).

\bibitem[{\citenamefont{{Santos} et~al.}(2007)}]{santos2007}
\bibinfo{author}{\bibfnamefont{D.}~\bibnamefont{{Santos}}}
  \bibnamefont{et~al.}, \bibinfo{journal}{J. Phys. Conf. Ser.}
  \textbf{\bibinfo{volume}{65}}, \bibinfo{pages}{012012}
  (\bibinfo{year}{2007}), \eprint{arXiv:astro-ph/0703310}.

\bibitem[{\citenamefont{{Nishimura} et~al.}(2008)}]{nishimura2008}
\bibinfo{author}{\bibfnamefont{H.}~\bibnamefont{{Nishimura}}}
  \bibnamefont{et~al.}, \bibinfo{journal}{J. Phys. Conf. Ser.}
  \textbf{\bibinfo{volume}{120}}, \bibinfo{pages}{042025}
  (\bibinfo{year}{2008}).

\bibitem[{\citenamefont{{Sciolla} et~al.}(2008)}]{sciolla2008}
\bibinfo{author}{\bibfnamefont{G.}~\bibnamefont{{Sciolla}}}
  \bibnamefont{et~al.} (\bibinfo{year}{2008}), \bibinfo{note}{arXiv:0805.2431}.

\bibitem[{\citenamefont{{Amram} et~al.}(1999)}]{amram1999}
\bibinfo{author}{\bibfnamefont{P.}~\bibnamefont{{Amram}}} \bibnamefont{et~al.},
  \bibinfo{journal}{Nucl. Phys. B Proc. Suppl.} \textbf{\bibinfo{volume}{75}},
  \bibinfo{pages}{415} (\bibinfo{year}{1999}).

\bibitem[{\citenamefont{{Hill} et~al.}(2006)}]{hill2006}
\bibinfo{author}{\bibfnamefont{G.~C.} \bibnamefont{{Hill}}}
  \bibnamefont{et~al.} (\bibinfo{year}{2006}), \eprint{arXiv:astro-ph/0611773}.

\bibitem[{\citenamefont{{Encyclop{\ae}dia Britannica}}(1994-1999)}]{earth1}
\bibinfo{author}{\bibnamefont{{Encyclop{\ae}dia Britannica}}},
  \emph{\bibinfo{title}{{The Earth: Its Properties, Composition, and
  Structure}}} (\bibinfo{publisher}{Britannica CD, Version 99. Encyclop{\ae}dia
  Britannica, Inc.}, \bibinfo{year}{1994-1999}).

\bibitem[{\citenamefont{{McDonough}}(2003)}]{earth2}
\bibinfo{author}{\bibfnamefont{W.~F.} \bibnamefont{{McDonough}}},
  \emph{\bibinfo{title}{{Treatise on Geochemistry}}}, vol.~\bibinfo{volume}{2}
  (\bibinfo{publisher}{Amsterdam, Elsevier}, \bibinfo{year}{2003}).

\bibitem[{\citenamefont{{Gondolo} et~al.}(2004)}]{gondolo2004}
\bibinfo{author}{\bibfnamefont{P.}~\bibnamefont{{Gondolo}}}
  \bibnamefont{et~al.}, \bibinfo{journal}{JCAP} \textbf{\bibinfo{volume}{7}},
  \bibinfo{pages}{8} (\bibinfo{year}{2004}), \eprint{arXiv:astro-ph/0406204}.

\bibitem[{\citenamefont{{The IceCube Collaboration}}(2001)}]{icecube2001}
\bibinfo{author}{\bibnamefont{{The IceCube Collaboration}}}
  (\bibinfo{year}{2001}), \bibinfo{note}{http://www.icecube.wisc.edu/\newline
  science/publications/pdd/pdd.pdf}.

\bibitem[{\citenamefont{{Akerib} et~al.}(2006{\natexlab{b}})}]{akerib2006}
\bibinfo{author}{\bibfnamefont{D.~S.} \bibnamefont{{Akerib}}}
  \bibnamefont{et~al.}, \bibinfo{journal}{Phys. Rev. Lett.}
  \textbf{\bibinfo{volume}{96}}, \bibinfo{pages}{011302}
  (\bibinfo{year}{2006}{\natexlab{b}}).

\bibitem[{\citenamefont{{Bergstr{\"o}m}
  et~al.}(1998{\natexlab{b}})\citenamefont{{Bergstr{\"o}m}, {Edsj{\"o}}, and
  {Gondolo}}}]{bergstrom1998}
\bibinfo{author}{\bibfnamefont{L.}~\bibnamefont{{Bergstr{\"o}m}}},
  \bibinfo{author}{\bibfnamefont{J.}~\bibnamefont{{Edsj{\"o}}}},
  \bibnamefont{and}
  \bibinfo{author}{\bibfnamefont{P.}~\bibnamefont{{Gondolo}}},
  \bibinfo{journal}{Phys. Rev. D} \textbf{\bibinfo{volume}{58}},
  \bibinfo{pages}{103519} (\bibinfo{year}{1998}{\natexlab{b}}),
  \eprint{arXiv:hep-ph/9806293}.

\bibitem[{\citenamefont{{Kerr} and {Lynden-Bell}}(1986)}]{kerr1986}
\bibinfo{author}{\bibfnamefont{F.~J.} \bibnamefont{{Kerr}}} \bibnamefont{and}
  \bibinfo{author}{\bibfnamefont{D.}~\bibnamefont{{Lynden-Bell}}},
  \bibinfo{journal}{Mon. Not. Roy. Astron. Soc.}
  \textbf{\bibinfo{volume}{221}}, \bibinfo{pages}{1023} (\bibinfo{year}{1986}).

\bibitem[{\citenamefont{{Reid}}(2009)}]{reid2009}
\bibinfo{author}{\bibfnamefont{M.~J.} \bibnamefont{{Reid}}}, in
  \emph{\bibinfo{booktitle}{American Astronomical Society Meeting Abstracts}}
  (\bibinfo{year}{2009}), vol. \bibinfo{volume}{213} of
  \emph{\bibinfo{series}{American Astronomical Society Meeting Abstracts}}, pp.
  \bibinfo{pages}{325.02--+}.

\bibitem[{\citenamefont{{Dehnen} and {Binney}}(1998)}]{dehnen1998}
\bibinfo{author}{\bibfnamefont{W.}~\bibnamefont{{Dehnen}}} \bibnamefont{and}
  \bibinfo{author}{\bibfnamefont{J.~J.} \bibnamefont{{Binney}}},
  \bibinfo{journal}{Mon. Not. Roy. Astron. Soc.}
  \textbf{\bibinfo{volume}{298}}, \bibinfo{pages}{387} (\bibinfo{year}{1998}),
  \eprint{arXiv:astro-ph/9710077}.

\bibitem[{\citenamefont{{Navarro} et~al.}(1997)\citenamefont{{Navarro},
  {Frenk}, and {White}}}]{navarro1997}
\bibinfo{author}{\bibfnamefont{J.~F.} \bibnamefont{{Navarro}}},
  \bibinfo{author}{\bibfnamefont{C.~S.} \bibnamefont{{Frenk}}},
  \bibnamefont{and} \bibinfo{author}{\bibfnamefont{S.~D.~M.}
  \bibnamefont{{White}}}, \bibinfo{journal}{\apj}
  \textbf{\bibinfo{volume}{490}}, \bibinfo{pages}{493} (\bibinfo{year}{1997}),
  \eprint{astro-ph/9611107}.

\bibitem[{\citenamefont{{Moore} et~al.}(2001)\citenamefont{{Moore},
  {Calc{\'a}neo-Rold{\'a}n}, {Stadel}, {Quinn}, {Lake}, {Ghigna}, and
  {Governato}}}]{moore2001}
\bibinfo{author}{\bibfnamefont{B.}~\bibnamefont{{Moore}}},
  \bibinfo{author}{\bibfnamefont{C.}~\bibnamefont{{Calc{\'a}neo-Rold{\'a}n}}},
  \bibinfo{author}{\bibfnamefont{J.}~\bibnamefont{{Stadel}}},
  \bibinfo{author}{\bibfnamefont{T.}~\bibnamefont{{Quinn}}},
  \bibinfo{author}{\bibfnamefont{G.}~\bibnamefont{{Lake}}},
  \bibinfo{author}{\bibfnamefont{S.}~\bibnamefont{{Ghigna}}}, \bibnamefont{and}
  \bibinfo{author}{\bibfnamefont{F.}~\bibnamefont{{Governato}}},
  \bibinfo{journal}{Phys. Rev. D} \textbf{\bibinfo{volume}{64}},
  \bibinfo{pages}{063508} (\bibinfo{year}{2001}),
  \eprint{arXiv:astro-ph/0106271}.

\bibitem[{\citenamefont{{Peter}}(2008)}]{peter2008}
\bibinfo{author}{\bibfnamefont{A.~H.~G.} \bibnamefont{{Peter}}}, Ph.D. thesis,
  \bibinfo{school}{Princeton University} (\bibinfo{year}{2008}).

\bibitem[{\citenamefont{{Peter}}(2009{\natexlab{a}})}]{peter2009a}
\bibinfo{author}{\bibfnamefont{A.~.H.~G.} \bibnamefont{{Peter}}}, 
  (\bibinfo{year}{2009}{\natexlab{a}}), \bibinfo{note}{arXiv:0902.1344}.

\bibitem[{\citenamefont{{Peter}}(2009{\natexlab{b}})}]{peter2009b}
\bibinfo{author}{\bibfnamefont{A.~.H.~G.} \bibnamefont{{Peter}}}, 
  (\bibinfo{year}{2009}{\natexlab{b}}), \bibinfo{note}{arXiv:0902.1347}.

\bibitem[{\citenamefont{{Vogelsberger}
  et~al.}(2008)\citenamefont{{Vogelsberger}, {Helmi}, {Springel}, {White},
  {Wang}, {Frenk}, {Jenkins}, {Ludlow}, and {Navarro}}}]{vogelsberger2008}
\bibinfo{author}{\bibfnamefont{M.}~\bibnamefont{{Vogelsberger}}},
  \bibinfo{author}{\bibfnamefont{A.}~\bibnamefont{{Helmi}}},
  \bibinfo{author}{\bibfnamefont{V.}~\bibnamefont{{Springel}}},
  \bibinfo{author}{\bibfnamefont{S.~D.~M.} \bibnamefont{{White}}},
  \bibinfo{author}{\bibfnamefont{J.}~\bibnamefont{{Wang}}},
  \bibinfo{author}{\bibfnamefont{C.~S.} \bibnamefont{{Frenk}}},
  \bibinfo{author}{\bibfnamefont{A.}~\bibnamefont{{Jenkins}}},
  \bibinfo{author}{\bibfnamefont{A.}~\bibnamefont{{Ludlow}}}, \bibnamefont{and}
  \bibinfo{author}{\bibfnamefont{J.~F.} \bibnamefont{{Navarro}}},
  \bibinfo{journal}{ArXiv e-prints}  (\bibinfo{year}{2008}),
  \eprint{arXiv:0812.0362}.

\bibitem[{\citenamefont{{Kuijken} and {Gilmore}}(1989)}]{kuijken1989b}
\bibinfo{author}{\bibfnamefont{K.}~\bibnamefont{{Kuijken}}} \bibnamefont{and}
  \bibinfo{author}{\bibfnamefont{G.}~\bibnamefont{{Gilmore}}},
  \bibinfo{journal}{Mon. Not. Roy. Astron. Soc.}
  \textbf{\bibinfo{volume}{239}}, \bibinfo{pages}{651} (\bibinfo{year}{1989}).

\bibitem[{\citenamefont{{Holmberg} and {Flynn}}(2000)}]{holmberg2000}
\bibinfo{author}{\bibfnamefont{J.}~\bibnamefont{{Holmberg}}} \bibnamefont{and}
  \bibinfo{author}{\bibfnamefont{C.}~\bibnamefont{{Flynn}}},
  \bibinfo{journal}{Mon. Not. Roy. Astron. Soc.}
  \textbf{\bibinfo{volume}{313}}, \bibinfo{pages}{209} (\bibinfo{year}{2000}),
  \eprint{arXiv:astro-ph/9812404}.

\bibitem[{\citenamefont{{Diemand} et~al.}(2004)\citenamefont{{Diemand},
  {Moore}, and {Stadel}}}]{diemand2004}
\bibinfo{author}{\bibfnamefont{J.}~\bibnamefont{{Diemand}}},
  \bibinfo{author}{\bibfnamefont{B.}~\bibnamefont{{Moore}}}, \bibnamefont{and}
  \bibinfo{author}{\bibfnamefont{J.}~\bibnamefont{{Stadel}}},
  \bibinfo{journal}{Mon. Not. Roy. Astron. Soc.}
  \textbf{\bibinfo{volume}{352}}, \bibinfo{pages}{535} (\bibinfo{year}{2004}),
  \eprint{arXiv:astro-ph/0402160}.

\bibitem[{\citenamefont{{Faltenbacher} and {Diemand}}(2006)}]{faltenbacher2006}
\bibinfo{author}{\bibfnamefont{A.}~\bibnamefont{{Faltenbacher}}}
  \bibnamefont{and}
  \bibinfo{author}{\bibfnamefont{J.}~\bibnamefont{{Diemand}}},
  \bibinfo{journal}{Mon. Not. Roy. Astron. Soc.}
  \textbf{\bibinfo{volume}{369}}, \bibinfo{pages}{1698} (\bibinfo{year}{2006}),
  \eprint{arXiv:astro-ph/0602197}.

\bibitem[{\citenamefont{{Lake}}(1989)}]{lake1989}
\bibinfo{author}{\bibfnamefont{G.}~\bibnamefont{{Lake}}},
  \bibinfo{journal}{Astron. J.} \textbf{\bibinfo{volume}{98}},
  \bibinfo{pages}{1554} (\bibinfo{year}{1989}).

\bibitem[{\citenamefont{{Read} et~al.}(2009)\citenamefont{{Read}, {Debattista},
  {Agertz}, {Mayer}, {Brooks}, {Governato}, and {Lake}}}]{read2009}
\bibinfo{author}{\bibfnamefont{J.~I.} \bibnamefont{{Read}}},
  \bibinfo{author}{\bibfnamefont{V.}~\bibnamefont{{Debattista}}},
  \bibinfo{author}{\bibfnamefont{O.}~\bibnamefont{{Agertz}}},
  \bibinfo{author}{\bibfnamefont{L.}~\bibnamefont{{Mayer}}},
  \bibinfo{author}{\bibfnamefont{A.~M.} \bibnamefont{{Brooks}}},
  \bibinfo{author}{\bibfnamefont{F.}~\bibnamefont{{Governato}}},
  \bibnamefont{and} \bibinfo{author}{\bibfnamefont{G.}~\bibnamefont{{Lake}}}
  (\bibinfo{year}{2009}), \eprint{arXiv:0901.2938}.

\bibitem[{\citenamefont{{Bruch} et~al.}(2009)\citenamefont{{Bruch}, {Peter},
  {Read}, {Baudis}, and {Lake}}}]{bruch2009}
\bibinfo{author}{\bibfnamefont{T.}~\bibnamefont{{Bruch}}},
  \bibinfo{author}{\bibfnamefont{A.~H.~G.} \bibnamefont{{Peter}}},
  \bibinfo{author}{\bibfnamefont{J.~I.} \bibnamefont{{Read}}},
  \bibinfo{author}{\bibfnamefont{L.}~\bibnamefont{{Baudis}}}, \bibnamefont{and}
  \bibinfo{author}{\bibfnamefont{G.}~\bibnamefont{{Lake}}}
  (\bibinfo{year}{2009}), \bibinfo{note}{in prep}.

\end{thebibliography}

\end{document}